\documentclass[preprint,11pt]{aastex}

\usepackage[utf8]{inputenc}
\usepackage{amsmath}
\usepackage{makecell}
\usepackage{tabularx}
\usepackage{graphicx}
\usepackage{fancyhdr}
\usepackage{lipsum}
\usepackage{booktabs, makecell, tabularx}
\usepackage{rotating}
\usepackage[export]{adjustbox}
\usepackage[british]{babel}
\usepackage{hhline}
\usepackage{multirow}
\usepackage[caption=false]{subfig}

\newcommand{\be}{\begin{equation}}
\newcommand{\ee}{\end{equation}}
\newcommand{\nn}{\mbox{} \nonumber \\ \mbox{} }
\newcommand{\ba}{\begin{eqnarray}}
\newcommand{\ea}{\end{eqnarray}}
\newcommand{\om}{\omega}
\newcommand{\Alfven}{Alfv\'{e}n }

\newcommand{\E}{{\bf E}}
\newcommand{\B}{{\bf B}}
\newcommand{\J}{{\bf J}}
\renewcommand{\v}{{\bf v}}
\renewcommand{\k}{{\bf k}}
\renewcommand{\div}{{\rm \,div\,}}

\newcommand\eg{{\it{e.g.}}}

\newcommand{\Bf}{{magnetic field}}
\newcommand{\Bfs}{{magnetic fields}}
\newcommand{\Ef}{{electric  field}}

\newcommand{\NS}{neutron star}
\newcommand{\NSs}{{neutron stars}}
\newcommand{\EM}{electromagnetic}
\newcommand{\BH}{{black hole}}
\newcommand{\BHs}{{black holes}}
\newcommand{\Sc}{Schwarzschild}
\newcommand{\ms}{magnetosphere}
\newcommand{\mss}{magnetospheres}

\newcommand{\LC}{light cylinder}
\newcommand{\Lf}{Lorentz factor}
\newcommand{\Lfs}{Lorentz factors}

\begin{document}

\title {Jump-starting  relativistic  flows, and the M87 jet}

\author{Maxim Lyutikov, Ahmad Ibrahim\\
Department of Physics and Astronomy, Purdue University, \\
 525 Northwestern Avenue,
West Lafayette, IN
47907-2036 }

\begin{abstract} 
We point out the dominant  importance of plasma injection effects for  relativistic winds from  pulsars and black holes. We demonstrate  that  outside the \LC\ the  magnetically dominated outflows while sliding along the helical \Bf\  move in fact  nearly radially with very large Lorentz factors $\gamma_0 \gg 1 $,  imprinted into the flow  during  pair production within the gaps. 
Only at larger distances,  $r \geq \gamma_0 (c/\Omega)$,
the MHD acceleration $\Gamma \propto r$  takes over. 
As a result, Blandford-Znajek (BZ) driven outflows  would produce spine-brightened images. The best-resolved case of the jet in M87 shows both bright edge-brightened features, as well as weaker spine-brightened feature. Only the spine-brightened component can be BZ-driven/originate from the BH's \ms.

\end {abstract}


  \section{Introduction}
  
  Acceleration of relativistic winds and jets is a classical problem in high energy astrophysics 
\citep[\eg][]{1969ApJ...158..727M,1970ApJ...160..971G,1977MNRAS.179..433B,1979ApJ...232...34B,1986A&A...162...32C,Krolik:1999,2006MNRAS.368.1561M,2008MNRAS.385L..28B,2019ARA&A..57..467B}. 
A standard approach involves solution of the MHD equation (analytical or numerical)  starting with a slowly moving plasma. Plasma is then accelerated by the corresponding pressure gradient, and collimated by magnetic hoop stresses \citep{1982MNRAS.199..883B}.

\begin{figure}[!htb]
\centering
  \includegraphics[width=0.6\linewidth]{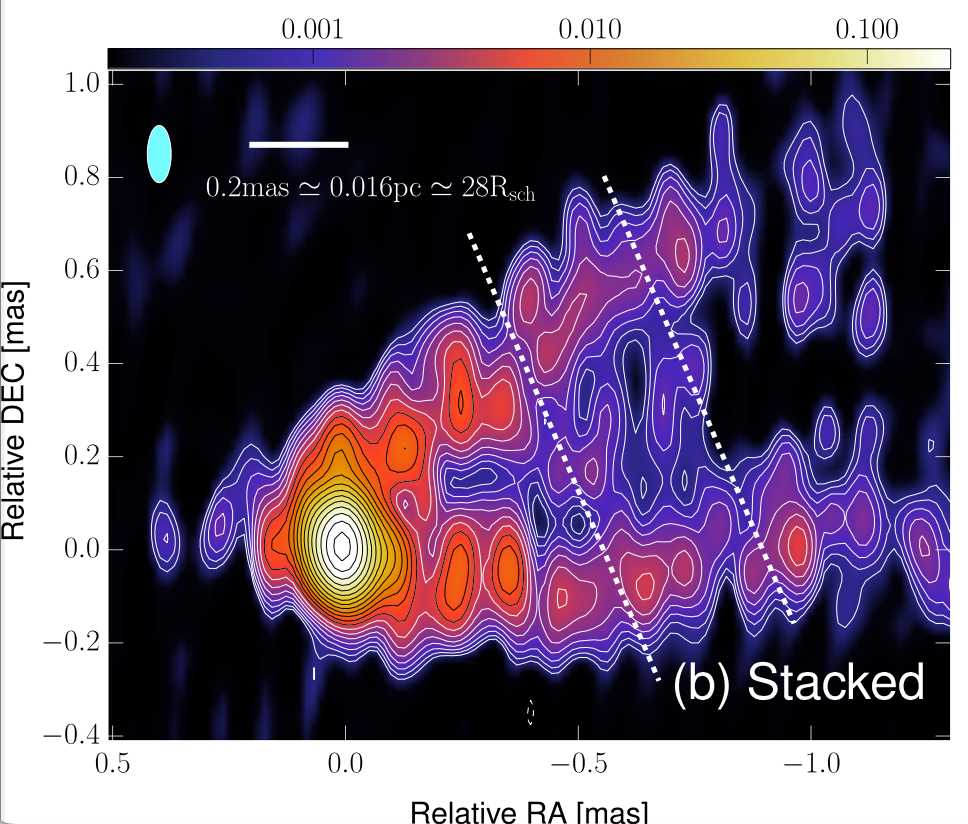}
  \includegraphics[width=0.7\linewidth]{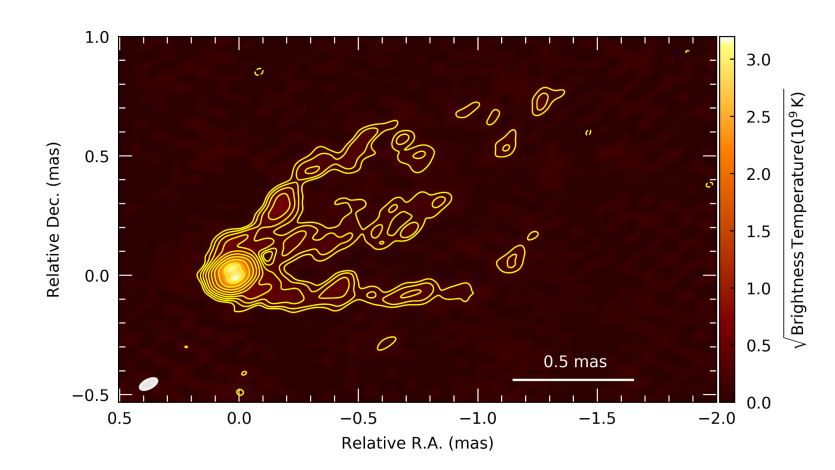}
\caption{ Top: Central part of jet in M87 at 86 GHz showing limb-brightened jet \citep{2018A&A...616A.188K}. {\bf } Bottom:   
 spine-brightened structure \citep{ lu2023ring}}.
\label{m87}
\end{figure}

{ 
Observations of the inner part of the jet in M87, down  to just 7 \Sc\ radii, 
shows complicated structure.  First, one observes 
 limb-brightened collimated  jet -  the jet accelerating
smoothly, with a parabolic profile  \cite{2013ApJ...775..118N,2018A&A...616A.188K,2019ARA&A..57..467B},  
Fig. \ref{m87}.

In addition to limb-brightened structures, \cite{lu2023ring} recently 
detected  a new feature  - spine-brightened jet.
}

{ 
In this work we aim to model  the  emission pattern expected in the  \cite{1977MNRAS.179..433B} model of jet acceleration and compare with the observation. 
The key new ingredient in our work is taking into account large initial (injection) velocities  parallel to the local \Bf. 
}

MHD models of acceleration \citep{BeskinBook,2009MNRAS.394.1182K,2017MNRAS.469.3840N}  take full account of plasma  velocity: both along and across \Bf. The corresponding analytical treatment, based on the relativistic Grad-Shafranov equation 
\citep{Grad1967,Shafranov1966,1973ApJ...182..951S,BeskinBook} is  completed, as it requires finding the initially unknown current distribution together with the solution  for the \Bf.

In the limit of highly magnetized plasma -  the force-free limit - the parallel velocity is not defined in principle.  But that does not mean it can be neglected. Plasma may be/is  streaming with large \Lfs\ along \Bfs. It seems this particular aspect was/is not considered previously. But it is highly important as we argue here.

Models of  gaps in \BH\ \mss\ \citep{1998ApJ...497..563H,1977MNRAS.179..433B,2000PhRvL..85..912L,2011ApJ...730..123L,2016A&A...593A...8P} generally predict that magnetospheric gaps, with thickness much smaller than the size of the \ms\ (the \LC) accelerate particles with {\Lf}s $\sim 10^3-10^4$; accelerated particles first IC scatter soft disk photons, this is followed by two-photon pair production, and the \EM\ cascade. Resulting  {\Lf}s are of similar values,  $\sim 10^3-10^4$. Lorentz factors up to $\sim 10^6$ are also possible \citep{2016A&A...593A...8P}.

In MHD   simulations, first, high magnetization is hard to achieve and, second, plasma is typically injected at rest \citep[\eg][]{1994PASJ...46..123T}.
 In the corresponding PIC simulations particles are typically injected at rest \citep[\eg][]{2014ApJ...795L..22C,2015ApJ...801L..19P,2020PhRvL.124n5101C,2022arXiv220902121H}. 

   Perhaps the closest  approach to the current one is 
   \cite{2000NCimB.115..795B}, where the importance of injection for the structure of the \ms\ and the corresponding  energy relations were discussed, quote "it is the pair creation region that plays the role of the energy source" \citep[][is also relevant]{1992SvA....36..642B}.

\section{The approach and the conclusion}


We start with a force-free solution and   add particle dynamics along the field kinematically, in the bead-on-wire approximation, neglecting its back reaction on the structure of the \Bf.
The bead-on-wire approach has     a clear advantage: for a given structure of the  \Bf\ the particle dynamics is easily calculated in {\it algebraic form} \citep[see also Section 7.2.6 of][]{2014MNRAS.445.2500G}. No 
 integration of the equations of motion is needed and no special conditions (\eg\ at \Alfven or fast surfaces)  appear.  The drawback is that it does not provide the full picture of what the  \Bf\ structure is: the structured the \ms\  should be  prescribed. Thus, our approach can be seen as the next term in expansion in  magnetization parameter $1/\sigma \ll 1$:
 force-free solution in the limit  $1/\sigma \to 0$ provide the structure of the \Bf, the  next term takes in the account particle dynamics in the prescribed \Bf.

 As a start, we assume that  flow lines and magnetic flux surfaces are  conical. (More complicated collimated flux surfaces behave similarly, see \S \ref{parabolicc}). In flat  metrics, 
 there is then analytical solution for the monopolar \Bf\ due to  \cite{1973ApJ...180L.133M} (it can be generalized to  \Sc\ case). We use it as a starting point: 
   \ba &&
   B_r = \frac{r_0^2}{r^2}  B_0
   \nn &&
   B_\phi = - \frac{r_0^2 \sin \theta \Omega }{r}  B_0 = E_\theta
   \nn &&
   \Vec{\beta}_{EM}=  \left\{\frac{r^2 \sin ^2\theta  \Omega ^2 }{1+r^2 \sin ^2\theta  \Omega ^2 },0,\frac{r
   \Omega  \sin (\theta )}{1+r^2 \sin ^2\theta  \Omega ^2 }\right\} 
  \nn && 
  \Gamma_{EM}= \sqrt{ 1+ r^2 \Omega ^2 \sin ^2(\theta )}
  \label{EM1} 
  \ea
  This analytical force-free solution  of the pulsar equation \citep{1973ApJ...182..951S,BeskinBook} passes smoothly through the \LC\  (\Alfven surface) $R_{LC} = c/(\sin \theta \Omega)$. Numerical models of the inner wind indicate that the structure of the \EM\ fields quickly approaches Michel's solution \cite{2018MNRAS.474.1526P}. \cite{2004MNRAS.350..427K} showed that monopolar geometry of \Bf\ lines is also a good approximation in case of  a \BH\ in external \Bf.

 \begin{figure}[h!]
\includegraphics[width=0.99\linewidth]{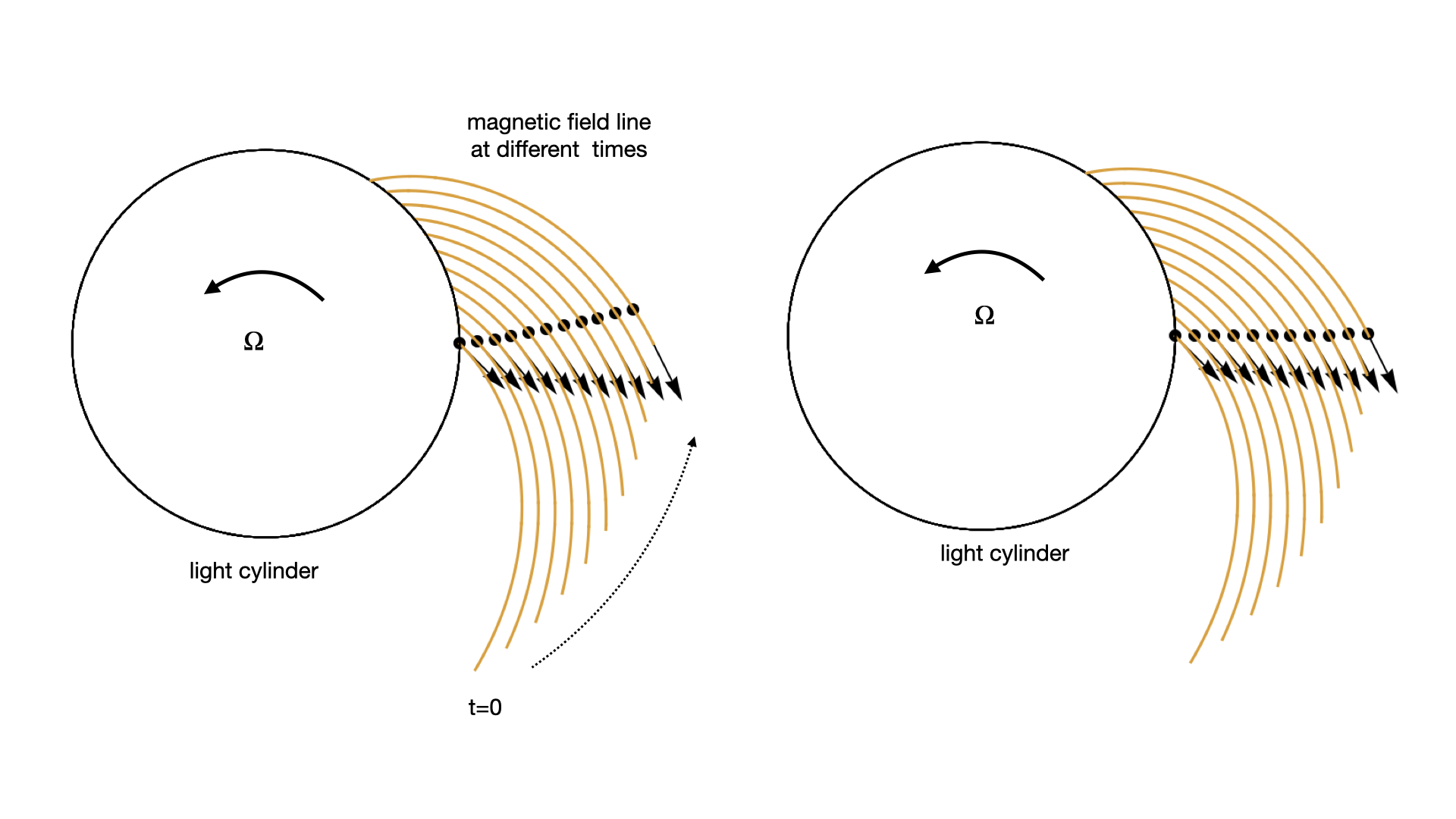}
    \caption{Particles' trajectories  calculated in Michel's field outside the \LC\ for \Lfs\  $\gamma_0=2, \, 10$, Eq, (\ref{trajjj}). The circle is the \LC,  orange curves indicate the same \Bf\ line at consecutive movements, arrows are directions of particle velocity at each point, and black dots are the location of the particle.  These calculations  illustrate that for $\gamma_0 \gg 1 $ the trajectory is nearly radial.}
\label{geom00}
    \end{figure}
    
  As long as the force-free condition is satisfied (negligible inertial effects) arbitrary  initial motion of (charge neutral)  plasma can be added along the field. Conditions at the \LC\ remain unchanged. The total velocity is then the (relativistic) sum of the \EM\ velocity (\ref{EM1}) and the motion along the rotating \Bf.

  It turns out that a particle launched with large \Lf\ along the rotating magnetic spiral (in a bead-on-wire approximation) moves nearly radially, Fig. \ref{geom00}.  This has been a known effect from numerical analysis \citep{2020MNRAS.491.5579C}, and was recently  analytically discussed by \cite{2022ApJ...933L...6L}.
   The key point is that the azimuthal  motion along the spiral is nearly compensated by the motion of the spiral itself.
  In this paper we generalize the results of \cite{2022ApJ...933L...6L}  for  particle dynamics in rotating winds to the curved space of \Sc\ and Kerr \BHs.

\begin{figure}[h!]
\includegraphics[width=0.99\linewidth]{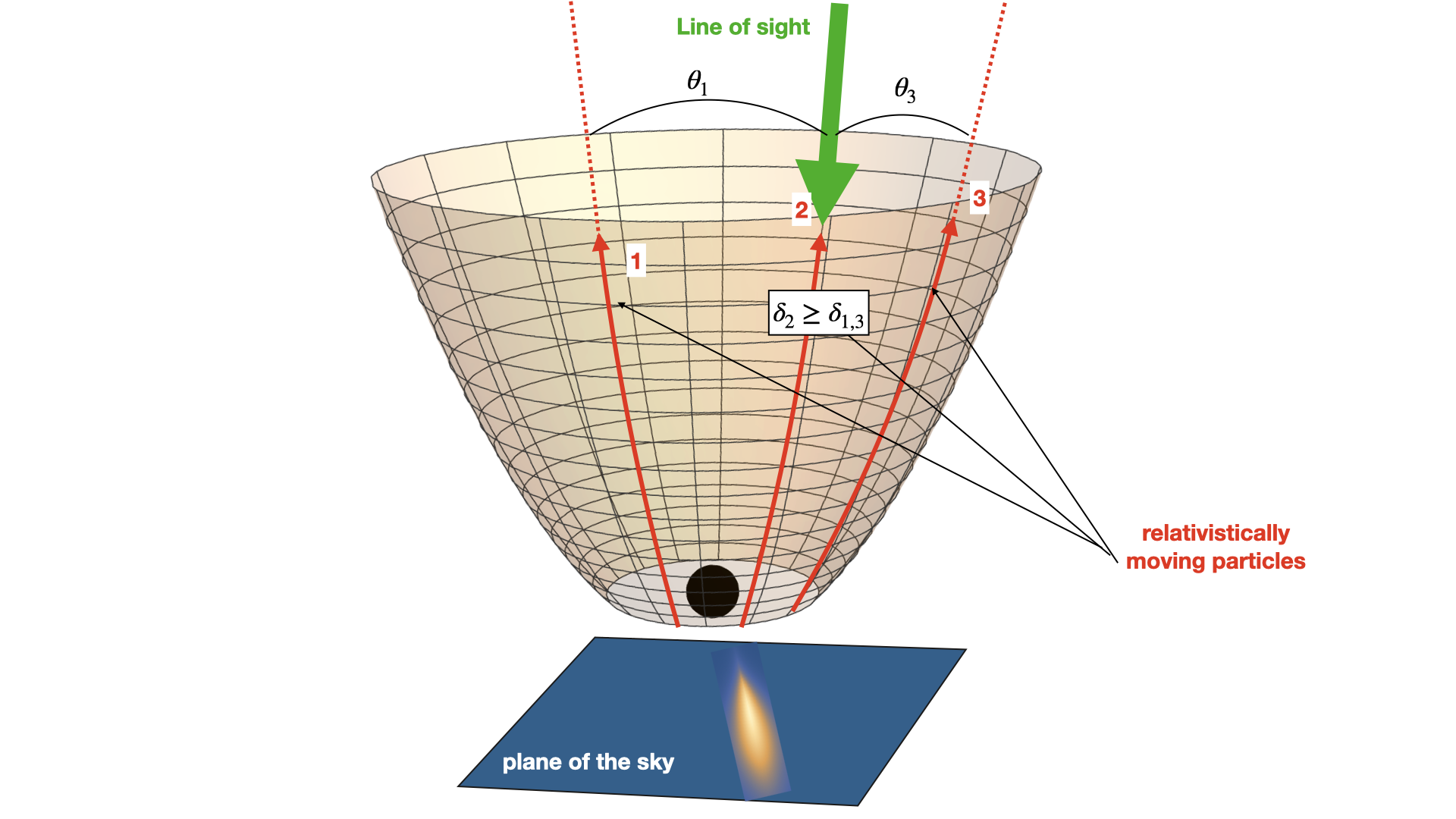}
    \caption{Graphic explanation why BZ-produced jets are expected to be spine-brightened. Shown are flux surfaces and  three different  particles' trajectories.  Emission produced by particle 2 has the smallest angle with respect to the line of sight,  largest Doppler factor,  and would result in  brightest emission pattern. }
\label{IC-BH-6}
    \end{figure}

The observed emission pattern from relativistically moving particles is dominated  by the Doppler factor  $\delta$. For axisymmetric jet the largest Doppler factor is along the flow lines that project to the spine of the image, Fig. \ref{IC-BH-6}.

 { Thus, relativistic flows emanating from within the \ms, driven by the BZ process,  are expected to produce spine-brightened image.  This is consistent with  observations of a weak central spine in 
 M87  jet, Fig. \ref{m87} bottom panel. 
On the other hand, the edge-brightened emission observed by \cite{2018A&A...616A.188K} is inconsistent   with the BZ process. The edge-brightened part 
 is not  centered on the \BH. This implies that a flow is only mildly relativistic. For example, estimates of the viewing angle $15^\circ -30^\circ$ degrees \citep{1996ApJ...467..597B} imply  bulk {\Lfs} $\sim 3-2$ (otherwise the emission would have been beamed away).

We then conclude that the limb-brightened and spine-brightened parts of the  M87 jet have different origin:  the spine-brightened part originates within the BH \ms\ driven by the Blandford-Znajek mechanism \citep{1977MNRAS.179..433B}. The edge-brightened component should have a different origin. For example, it can  be produced by the \cite{1982MNRAS.199..883B} mechanism, starting as a slow  accelerating flow from the accretion disk. 
}

\section{Particle motion along rotating spiral}

\subsection{General relations}.

The motion of a particle in bead-on-wire approximation can be derived algebraically for a given structure of \ms\ in the case of flat, \Sc\ and Kerr metrics.  As a basic case, we start  with particles moving in the equatorial plane in case of Kerr metric. We start with a particular case of Archimedian spiral with radial step equal $1/\Omega$. 

We use the machinery of General Relativity to treat particle motion in the rotating frame, in curved space-time.  
We start at the equatorial plane $\theta = \pi/2$ of Kerr metric. 
For flat and \Sc\  cases, where monopolar \Bf\ is an exact solution, the generalization to motion with  fixed arbitrary polar angle is  recovered later. 

{ 
The results of this section are further  re-derived/extended in the Appendices. 
In Appendix \ref{Lagrangianwithconstraint} we re-derive the corresponding relations using Lagrangian and Hamiltonian approaches for relativistic particle moving along a constrained path in flat space-time. In Appendix \ref{TheLagrangianApproachAppE} we use alternative formulation of the Lagrangian. In section \ref{Omega1} we allow for arbitrary radial step (effective, taking into account non-force-free effects of plasma loading). Finally, in the most mathematically advanced approach, in Appendix \ref{KERRBLACKHOLEAPPENDIXAHMAD} we discuss the most general case of field structure in Kerr metric. }
   
   Kerr metric in the equatorial plane  is defined by the metric tensor
   \ba &&
   g_{00} =  1 - \frac{2 M }{r }=  \alpha^2
   \nn &&
   g_{rr} = \frac{r^2}{\Delta }
   \nn &&
   g_{\phi \phi}  =a^2 M^2+r^2+  \frac{2 a^2 M^3}{r}\equiv \Delta_1
   \nn &&
   g_{0 \phi} =- \frac{2 a M^2}{r}
   \nn && 
   \Delta = a^2 M^2-2 M r+r^2
   \nn &&
   \Delta_1= a^2 M^2+r^2+  \frac{2 a^2 M^3}{r}\
   \label{Kerr}
   \ea
   where $a$ is the dimensionless Kerr parameter.

Consider  rotating  \BH\ \ms.  Since we use bead-on-wire approximation, we first need to find a structure of a given \Bf\ line. The rotating spiral is defined by two parameters: the angular velocity of rotation $\Omega$ and radial step 
$c/\Omega_1$. In the force-free approximation,   $\Omega=\Omega_1$    (see Appendix  \ref{Omega1} for a generalization $\Omega _ 1\neq \Omega$), the 
   fast  \EM\ mode propagates radially with   (setting $ds=0$  for null  trajectory and $d\phi=0$ for radial propagation) 
\be
\beta_F = \sqrt{\frac{\Delta  (r-2 M)}{r^3}} = \frac{\alpha \sqrt{\Delta}}{r} \to \alpha^2
\label{betaF}
\ee
(the latter limit is for \Sc\ case).

Angular velocity of the Lense-Thirring precession 
\be
\om_{LT} =  a
\frac{2  M^2}{r^3+ a^2 M^2 (2 M+r)}= \frac{2 a M^2}{r \Delta_1} = -\frac{g_{0\phi}}{g_{\phi\phi}}
\ee

Thus, the radial step of the spiral is given by
\ba &&
d \phi' =   \omega _{sp}  dr
\nn &&
\omega _{sp}  = \frac{\om_{LT} - \Omega} {\beta_F} = -\sqrt{\frac{g_{rr}}{g_{00}}}\left(\frac{g_{0\phi}}{g_{\phi\phi}}+\Omega\right) 
\label{spiral} 
\ea
(For propagation with sub-luminal velocity,  see Appendix \ref{Omega1}. In this case the radial step is smaller than $1/\Omega$. This can be due to back-reaction of plasma on \Bf\ lines.)

Next, in the Kerr metric, transferring to the frame  rotating with the \Bf\ (see Appendix \ref{limits} for corresponding limitations)


\be
d\phi \to d\phi'  + \Omega dt
\ee 
and imposing the spiral constraint (\ref{spiral}), 
we find metric coefficients in the rotating frame 
\ba && 
G_{00} = g_{00} {-} \left(  g_{\phi \phi} \Omega - \frac{ 4 a M^2}{r} \right) \Omega 
\nn &&
G_{rr} = 
 \frac{r^2}{\Delta } + g_{\phi \phi}   \omega _{sp}^2
 \nn &&
 G_{0 r} =  \left( -\frac{2 a M^2}{r} + {\Omega}\:g_{\phi \phi}  \right)  \omega _{sp} 
 \label{GG}
\ea
The contra-variant  metric is 
\ba && 
G^ {00} =  - \frac{G_{rr} }{\Delta_G}
\nn &&
G^{rr} =    \frac{G_{00} }{\Delta_G}
\nn &&
G^{0r} =   \frac{G_{0r } }{\Delta_G}
\nn && 
\Delta_G= G_{0 r} ^2 +  G_{00}  G_{rr}  
\ea

Using the Hamilton-Jacobi equation
\be
{ G^ {00} } (\partial _t S)^2 + 2 { G^{0 r} } (\partial _t S) (\partial _r S) +{ G^{rr} }  (\partial _r S) ^2=1
\ee

with a separation
\be
S= - \gamma_0  t + S_1(r),
\ee
we find
\be
{ G^{00} } \gamma_0^2 -  2 { G^{0r} }  \gamma_0 (\partial _r S) + { G^{rr} }  (\partial _r S) ^2=1
\ee
Thus,
\be
 (\partial _r S) = \frac{ G_{0r} }{ G_{00}}  \gamma_0 \pm \frac{ \sqrt{ \Delta_G(  \gamma_0^2  - G_{00} )}}{G_{00}}
 \ee
 This equation can be analytically integrated in flat space, giving the trajectory $r(t)$ \citep{2022ApJ...933L...6L}; see also alternative derivation in Appendix \ref{Lagrangianwithconstraint}, Eq. (\ref{trajjj}). But deriving $r(t)$  is an unnecessary yet complicated step, since we are not interested in the time dependence of the particle velocity, only in its coordinate dependence. 
 
 Differentiating with respect  to $\gamma_0$
 \be
\partial_{\gamma_0} (\partial _r S)=  \frac{ G_{0r} }{ G_{00}}  \pm   \frac{ \gamma_0}{G_{00}}   \sqrt{ \frac{ \Delta_G}{\gamma_0^2 - G_{00}} } 
\ee
Finally,
\be
\beta_r = \left( \partial_{\gamma_0} (\partial _r S)\right)^{-1}  =
\frac{G_{00}}  { \gamma_0 \sqrt{ (G_{00} G_{rr} + G_{0r}^2)/(\gamma_0^2 - G_{00} ) }   + G_{0r}}
\label{MAIN}
\ee

In equation (\ref{MAIN}), we  ignored the negative solution because it represents going in a not interesting direction. We explain this more while re-deriving equation (\ref{MAIN}) by various  Lagrangian approaches in Appendices  \ref{Lagrangianwithconstraint} and  \ref{TheLagrangianApproachAppE}. This solves the problem of particle dynamics in rotating  \ms\ in the bead-on-wire approximation.

One of the mathematical complications involves changing the sign of $G_{00}$ while crossing the \LC. 
However, the speed $\beta_r$ stays positive. Explicit forms are given below, \eg\ Eq. (\ref{brflat}), passes smoothly through the \LC. 

We point out that the Hamilton-Jacobi  approach allows one to find trajectory purely algebraically - no integration of the equation of motion is involved. 

Next, we give explicit relations for particular examples of flat, \Sc, and Kerr  spaces.

\subsection{Flat space $M=0$}
\begin{figure}[h!]
\includegraphics[width=0.99\linewidth]{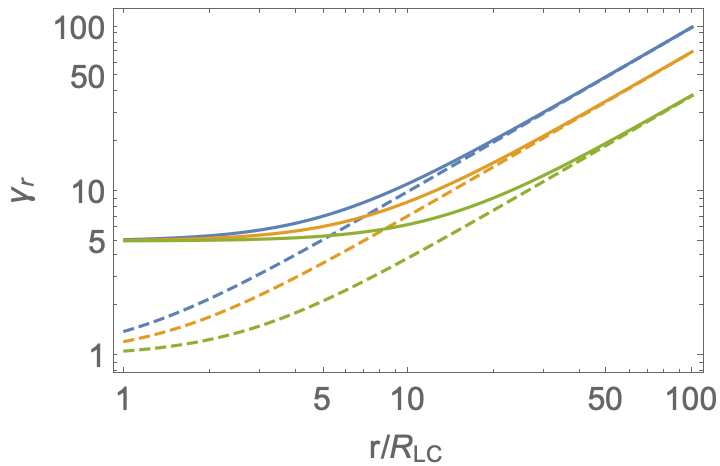}
    \caption{\Lfs\ corresponding to (\ref{gamma}) compared with Michels' solution $\gamma= \sqrt{1+ r^2 \Omega ^2 \sin^2 \theta}$ for $\gamma_0 =5$ and $\theta =\pi/8,\, \pi/4,\, \pi/2$ (bottom to top). Only at large distances $r \geq \gamma_0 R_{LC}$ MHD acceleration picks up.}
\label{compareMichel}
    \end{figure}

In the frame rotating with the spiral the  metric tensor is
\citep[See also][] {2022ApJ...933L...6L}
 \ba &&
   G_{00} =  1 - 
   \sin^2 \theta 
   r^2 \Omega^2
   \nn &&
   G_{0r} =
    -\sin^2 \theta
   { r^2 \Omega^2}
   \nn &&
   G_{rr} =  1 +
    \sin^2 \theta
{  r^2 \Omega^2}
   \label{Gflat}
   \ea
Gives Christoffel coefficients
\ba &&
\Gamma^0_{00} =   r^3 \sin^4 \theta  \Omega^4
\nn &&
\Gamma^0_{0r} =  - r  \sin^2 \theta \Omega^2 ( 1+ \sin^2 \theta r^2 \Omega^2)
\nn &&
\Gamma^0_{rr} =   r \sin^2 \theta \Omega^2 ( 2+ \sin^2 \theta r^2 \Omega^2)
\nn &&
\Gamma^r_{rr}= r \sin^2 \theta \Omega^2 ( 1+ \sin^2 \theta r^2 \Omega^2)
\nn &&
\Gamma^r_{0r}=  -r^3 \sin^4 \theta\Omega^4
\nn &&
\Gamma^r_{00}=-r \sin^2 \theta \Omega^2 ( 1- \sin^2 \theta r^2 \Omega^2)
\label{Christoffel}
\ea

\be
\beta_r =\frac{1-  \sin^2 \theta r^2 \Omega ^2}{\frac{\gamma _0}{\sqrt{\gamma _0^2+   \sin^2 \theta r^2 \Omega ^2-1}}-   \sin^2 \theta r^2 \Omega ^2}=
\left\{ 
 \begin{array}{cc}
 1-\frac{1}{2 \gamma _0^2} ,& \gamma_0 \gg r \Omega \sin \theta
 \\
 1- \frac{1}{  \sin^2 \theta r^2 \Omega ^2} + \frac{\gamma _0}{ \sin^3 \theta r^3 \Omega ^3}, & r \to \infty
 \\
 \frac{2 \gamma _0^2}{1+ 2 \gamma _0^2} \approx 1- \frac{1}{2 \gamma_0^2},  & r=1/\Omega
\end{array}
\right.
\label{brflat} 
\ee
The corresponding \Lf\ $\gamma=1/\sqrt{1-\beta_r^2 - \left(\sin \theta r \Omega(1-\beta_r) \right)^2}$ is
\be
\gamma = \frac{\gamma _0-  \sin^2 \theta r^2 \Omega ^2 \sqrt{\gamma _0^2+  \sin^2 \theta r^2 \Omega ^2-1}}{1-  \sin^2 \theta r^2 \Omega ^2}
=\left\{ 
 \begin{array}{cc}
\gamma _0+\frac{  \sin^2 \theta r^2 \Omega ^2}{2 \gamma _0} ,& \gamma_0 \gg r \Omega \sin \theta
 \\
 \sin \theta r \Omega +  \frac{1+ \gamma _0^2}{2 r   \sin \theta \Omega }, & r \to \infty\\
 \gamma _0  +\frac{1}{2 \gamma _0}  +\frac{  \sin \theta \Omega }{\gamma _0}  \left( r- \frac{1}{  \sin \theta\Omega} \right) , & r \to 1/(\sin \theta \Omega)
\end{array}
\right.
\label{gamma} 
\ee
see Fig. \ref{compareMichel}.

It may be verified using Lorentz transformations that the \Ef\ in the frame of the particle is zero.

The integration constant $\gamma_0$ physically corresponds to some value of the energy at some location.
With a change of parameter $\gamma_0$, so that $ \gamma_0 + 1/( {2 \gamma _0})  \to \gamma _{LC}$, the solution would correspond to initial condition $\gamma _{LC}$ on the \LC:


\be
\beta_r =\frac{1-  \sin^2 \theta r^2 \Omega ^2}{\frac{\gamma _{LC}+\sqrt{\gamma _{LC}^2-2}}{2
   \sqrt{\frac{1}{4} \left(\gamma _{LC}+\sqrt{\gamma
   _{LC}^2-2}\right){}^2+  \sin^2 \theta r^2 \Omega ^2-1}}-  \sin^2 \theta r^2 \Omega ^2}
   \label{gammach}
   \ee 
In what follows we skip this unnecessary redefinition.
 Numerically $\gamma_0$ is typically very  close to the energy of the particle crossing the outer \LC, see \eg\ (\ref{brflat}). Also, in  appendix (\ref{KERRBLACKHOLEAPPENDIXAHMAD}) we discuss the range of the allowed values for this constant of motion and we show that this constant behaves the same as the initial Lorentz factor of a particle at the outer light cylinder when $\gamma_0\gg 1$; hence, we give this constant the symbol $\gamma_0$.

   Importantly,  toroidal component of the velocity always remains small 
   \be
   \beta_\phi=  \sin \theta r \Omega   (1-\beta_r)  =
  \frac{\gamma _0-\sqrt{\gamma _0^2+r^2 \Omega ^2 \sin ^2(\theta )-1}}{\gamma _0-r^2 \Omega ^2 \sin
   ^2(\theta ) \sqrt{\gamma _0^2+r^2 \Omega ^2 \sin ^2(\theta )-1}} r \Omega  \sin (\theta )
    \label{Gamma3}
   \ee
its maximal value is reached at
$r \approx 1.27 \gamma_0 / \Omega$ and equals
\be
\beta_{\phi, max} \approx 0.3/\gamma_0
\label{betaphimax}
\ee

\begin{figure}[h!]
\includegraphics[width=0.99\linewidth]{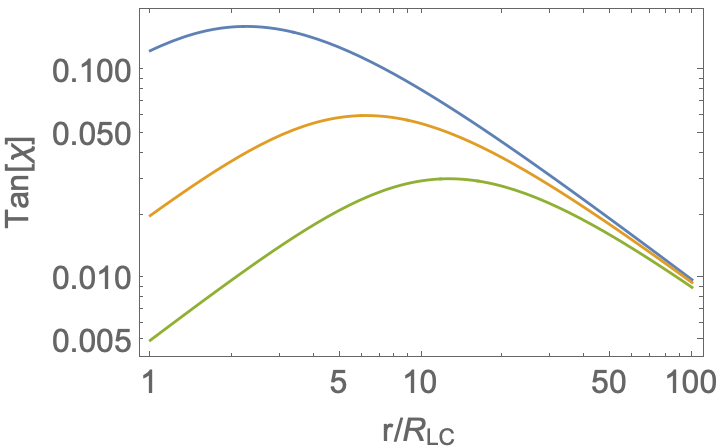}
    \caption{Angle of motion with respect to the radial direction, $\tan \chi = \beta_\phi/\beta_r$ for $\gamma_0 = 2,5, 10$ (top to bottom); flat space $M=0$. The motion is nearly radial.}
\label{pitchangle}
    \end{figure}

The angle of motion with respect to the radial direction, $\tan \chi = \beta_\phi/\beta_r$ always remains small, Fig. \ref{pitchangle}

    We also note simple relations in terms of proper time $\tau$ (for $\theta =\pi/2$).
    \ba && 
    \frac{d r}{d \tau} = \gamma \frac{d r}{d t} = 
    \sqrt{ \gamma_0^2  + r^2 \Omega^2 -1 } 
    \nn &&
    \frac{d^2 r}{d \tau^2}= r\Omega^2
    \label{d2tau}
    \ea

Eq. (\ref{d2tau}) has  a solution
\ba &&
  \Omega   r(\tau) = \gamma _0 \sinh (\tau  \Omega )+\cosh (\tau  \Omega )
  \nn &&
   r'(\tau) = \gamma _0 \cosh  (\tau  \Omega ) +  \sinh (\tau  \Omega )
  \ea
  (using $r(0)=1/\Omega$ and  $r'(0)=\gamma_0$).
In proper time the \Lf\ doubles approximately in  $\tau \sim 1/\Omega$. This result  is  consistent with doubling in observer time in  $t \sim \gamma_0 /\Omega$, at distance  $r \sim \gamma_0 R_{LC}$

\subsection{\Sc\ \BH, $a=0$}
  Metric tensor is now
   \ba &&
   G_{00} =  \alpha^2 - 
   \sin^2 \theta 
   r^2 \Omega^2= G^{rr} 
   \nn &&
   G_{0r} =-
    \sin^2 \theta
   \frac{ r^2 \Omega^2}{\alpha^2} = G^{0r} 
   \nn &&
   G_{rr} =  \alpha^{-2} +
    \sin^2 \theta
   \frac{  r^2 \Omega^2}{\alpha^4} = G^{00}
   \nn &&
   ||G|| = -1
   \label{GSc}
   \ea
(a factor of minus one is explicitly included in the definition of  the 0-0 component of the metric). 

The     Christoffel symbols evaluate to
\ba &&
\Gamma^0_{00} =-
\frac{\Omega ^2
   \sin ^2(\theta ) \left(M-r^3 \Omega ^2 \sin ^2(\theta )\right)}{\alpha ^2}
\nn &&
\Gamma^0_{0r} =
   -\frac{\left(r^3 \Omega ^2 \sin ^2(\theta
   )-M\right) \left(\alpha ^2+r^2 \Omega ^2 \sin ^2(\theta )\right)}{\alpha ^4 r^2}
   \nn &&
\Gamma^0_{rr} =
   \frac{\Omega ^2 \sin ^2(\theta ) \left(-5
   M+r^3 \Omega ^2 \sin ^2(\theta )+2 r\right)}{\alpha ^6}
\nn &&
\Gamma^r_{rr}= 
   \frac{\Omega ^2 \sin ^2(\theta ) \left(-3 M+r^3 \Omega ^2 \sin ^2(\theta )+r\right)}{\alpha
   ^4}+\frac{2M^2}{r^3\alpha^4}-\frac{M}{\alpha ^4 r^2}
\nn &&
\Gamma^r_{0r}= 
\frac{\Omega ^2 \sin ^2(\theta ) \left(M-r^3 \Omega ^2 \sin
   ^2(\theta )\right)}{\alpha ^2}
\nn &&
\Gamma^r_{00}=
   \frac{\left(r^3 \Omega ^2 \sin ^2(\theta )-M\right) \left(r^3
   \Omega ^2 \sin ^2(\theta )-\alpha ^2 r\right)}{r^3}
\label{ChristoffelSc}
\ea

    The radial velocity now is 
   \be
   \beta_r = \alpha^2 
   \frac{\alpha ^2-  \sin^2 \theta r^2 \Omega ^2}{\frac{\alpha ^2 \gamma _0}{\sqrt{\gamma _0^2+  \sin^2 \theta r^2 \Omega
   ^2 -\alpha ^2 }}- \sin^2 \theta r^2 \Omega ^2}
      =
\left\{ 
 \begin{array}{cc}
\alpha^2 \left(1 - \frac{\alpha^2}{ 2 \gamma_0^2}  \right),   & \gamma_0 \gg r \Omega \sin \theta
 \\
\alpha^2  \left( 1 - \frac{1}{ \sin^2 \theta r^2 \Omega ^2}+ \frac{\gamma_0 }{\alpha^2 \sin^3 \theta r^3 \Omega ^3} \right) , & r \to \infty
\end{array}
\right.
\label{betarSc}
\ee
Factors of $\alpha^2$ in front  are just relativistic coordinate  time dilation for a particle moving in gravitational  field. 

\subsection{Kerr \BH}
In the equatorial plane we find 
\be 
\beta_r = 
\frac{\alpha  \sqrt{\Delta } \Delta _1 \left(r \left(\alpha ^2-\Delta _1 \Omega ^2\right)-4 a M^2
   \Omega \right)}{\gamma _0 \sqrt{\frac{r \left(4 a^2 \Delta _1 M^4 r^2 \left(\alpha ^2+\Delta _1
   \Omega ^2\right)-16 a^3 \Delta _1 M^6 r \Omega +16 a^4 M^8-8 a \alpha ^2 \Delta _1^2 M^2 r^3
   \Omega +\alpha ^4 \Delta _1^2 r^4\right)}{4 a M^2 \Omega +r \left(-\alpha ^2+\gamma _0^2+\Delta
   _1 \Omega ^2\right)}}+4 a^2 M^4-\Delta _1^2 r^2 \Omega ^2}
 \ee
 For $ r \to \infty$ we find
\be
\beta_r =\alpha ^2-  \frac{1-a^2 M^2 \Omega ^2/2}{ r^2 \Omega ^2}+\frac{\gamma_0+ 4 M \Omega  (1- a M \Omega /2)}{r^3 \Omega ^3} 
   \ee

      It is understood that the solutions above involve both the kinematic  effects of time-dilation (\eg\ factor of $\alpha$ in front for the \Sc\ case), as well as effects of centrifugal acceleration.

In conclusion,  in all cases a particle moves radially with $\gamma \approx \gamma_0$ until  $r  \Omega \sim \gamma_0$. After that the wind acceleration takes over with 
$\gamma \sim r  \Omega $.

   \subsection{Spiral with arbitrary  radial step}
   \label {Omega1}
   The radial step of a magnetic spiral (denoted below $\equiv c/\Omega_1$) may be different from $c/\Omega$, \eg, the field may be affected by plasma inertia. For example,
 it is expected that inertial effects will make the spiral more tightly bound, larger $d \phi' $ for a given $dr$, hence  $\Omega_1 \geq \Omega$. 
   Equivalently, if the fast mode propagates with $\beta_s \leq 1$, in (\ref{betaF}),
   \ba &&
   \beta_F  \to \beta_s  \beta_F
   \nn &&
   \Omega_1 = \frac{ \Omega}{\beta_s} \geq  \Omega
   \ea

{ Most importantly, the radial velocity of  fast mode's propagation may change with radius depending on local plasma parameters. In this case the shape of the spiral is non-Archimedean. In our notations this implies radial dependence of the radial step: $\Omega_1(r)$. }
   
  Corresponding relations are fairly compact  in  \Sc\ metric. Using
   \ba &&
   d\phi \to d \phi' - \Omega dt
   \nn && 
    d \phi'  \to - \frac{\Omega_1}{\alpha^2} dr
    \ea
    (instead of corresponding relations (\ref{spiral}) with $\Omega_1 = \Omega$),
    we find
    \ba\label{PhysicallyRealizableStep} &&
    \beta_r = 
  \frac{\alpha ^2 \left(\alpha ^2-r^2 \Omega ^2\right)}{\frac{\alpha  \gamma _0 \sqrt{\alpha ^2+r^2
   \left(\Omega _1^2-\Omega ^2\right)}}{\sqrt{-\alpha ^2+\gamma _0^2+r^2 \Omega ^2}}-r^2 \Omega 
   \Omega _1}
   =
   \left\{ 
 \begin{array}{cc}
\alpha ^2 \left(\frac{\alpha ^2-r^2 \Omega ^2}{\alpha  \sqrt{\alpha ^2+r^2 \left(\Omega _1^2-\Omega
   ^2\right)}-r^2 \Omega  \Omega _1}
 \right),   & \gamma_0 \gg r \Omega
 \\
\alpha ^2 \left(\frac{\Omega }{\Omega _1}
-\frac{
    \left(   \Omega _1- \gamma _0 \sqrt{\Omega _1^2-\Omega ^2}\right)}{\alpha^2 r^2 \Omega _1^2 \Omega
   }
    \right) ,  & r \to \infty
\end{array}
\right.
   \ea

   For flat space
   \be
     \beta_r = 
 \frac{1-r^2 \Omega ^2}{\frac{\gamma _0 \sqrt{1+r^2 \left(\Omega _1^2-\Omega ^2\right)}}{\sqrt{\gamma
   _0^2+r^2 \Omega ^2-1}}-r^2 \Omega  \Omega _1}
    =
   \left\{ 
 \begin{array}{cc}
 \frac{1-r^2 \Omega ^2}{  \sqrt{1+r^2 \left(\Omega _1^2-\Omega
   ^2\right)}-r^2 \Omega  \Omega _1}
   \left( 1-\frac{\left(1-r^2 \Omega ^2\right) \sqrt{r^2 \left(\Omega _1^2-\Omega ^2\right)+1}}{2 \gamma _0^2
   \left(\sqrt{r^2 \left(\Omega _1^2-\Omega ^2\right)+1}-r^2 \Omega  \Omega _1\right)}\right) ,   & \gamma_0 \gg \Omega r

 \\ 
   \frac{\Omega }{\Omega _1}
-\frac{
    \left(   \Omega _1- \gamma _0 \sqrt{\Omega _1^2-\Omega ^2}\right)}{r^2 \Omega _1^2 \Omega},  & r \to \infty
    \end{array}
\right.
      \ee
 { Importantly, in the above relations the function $\Omega_1(r)$ is arbitrary, limited only by the condition $\Omega_1 \geq \Omega$. (So that a radial step per rotation can be smaller or equal to the \LC.

 For  given  $\Omega$ and $\Omega_1$ the toroidal velocity is 
 \be
 \beta_\phi = r \Omega \left( 1- \frac{\Omega_1} {\Omega} \beta_r \right)
 \ee
 Note that here $\Omega_1$  may depend on radius $r$. 
 
  On physical grounds we expect $\Omega_1 \approx \Omega$. In other words, fast mode is nearly relativistic and plasma is nearly force-free. This is the intrisic assumption of the model. For example, if $\Omega_1 = (1+\delta_\Omega)\Omega$, $\delta_\Omega\ll 1$, then
  \be
 \beta_\phi = r \Omega(1-\beta_r + \delta _\Omega)
 \ee
 Thus, for Archemedian spiral with the radial step $1/\Omega$, a particle slightly overtakes the rotation pattern with the rate $\sim r \Omega(1-\beta_r)\ll r \Omega$, while for $1-\beta_r \ll \delta _\Omega $  a particle lags behind.

Thus,  azimuthal velocity is small right from the \LC.  Maximal toroidal velocity is  $\sim 1/\gamma_0$, or $\sim \delta _\Omega$. }

\section{Numerical test}
\label{numerical}


We have developed a Boris-based  pusher \citep{boris_69,birdsall}. We verify the analytical results with direct integration, as we discuss below.

For \cite{1973ApJ...180L.133M} solution,  consider a particle that in the wind frame (boosted by $\beta_{EM}$ from the lab frame) moves with \Lf\ $\gamma_0'$ (prime indicates \Lf\ measured in the flow frame).  Using \EM\ velocity (\ref{EM1}) and  Lorentz transformations, we find the  momentum in the lab frame $p_\parallel$
\be
p_{\parallel}=    \left\{\frac{\beta _0' \gamma _0'+\left(\gamma _0'-1\right) r^2 \sin ^2\theta  \Omega ^2 }{\sqrt{1+r^2 \sin ^2\theta  \Omega ^2 }},0,\frac{\left(\sqrt{1-\beta
   _0'}-\sqrt{\beta _0'+1}\right) r \Omega  \sin (\theta )}{\sqrt{\beta _0'+1} \sqrt{1+ r^2  \Omega ^2 \sin ^2(\theta )}}\right\}
   \ee
 The momentum $p_\parallel$ is that of a particle that is sliding  along the local \Bf\ with \Lf\ $\gamma_0'$ as measured in the frame associated with $\beta_{EM}$ (where \Ef\ is zero). This is not a radial dependence, only a transformation at a given radius. 
 
   The total \Lf\ is 
   \be
   \gamma = \gamma_0'  \sqrt{ 1+   \sin^2 \theta r^2 \Omega^2 } 
   = \sqrt{2}  \gamma_0'
   \label{gamma22}
   \ee
    a combination of parallel  motion and orthogonal E-cross-B drift. The final relation in (\ref{gamma22}) applies to the \LC.
   
   In Fig.4 we compare analytical results (\ref{gamma}), and numerical integration - they are in excellent agreement.
   
    \begin{figure}[!htb]
\includegraphics[width=0.99\linewidth]{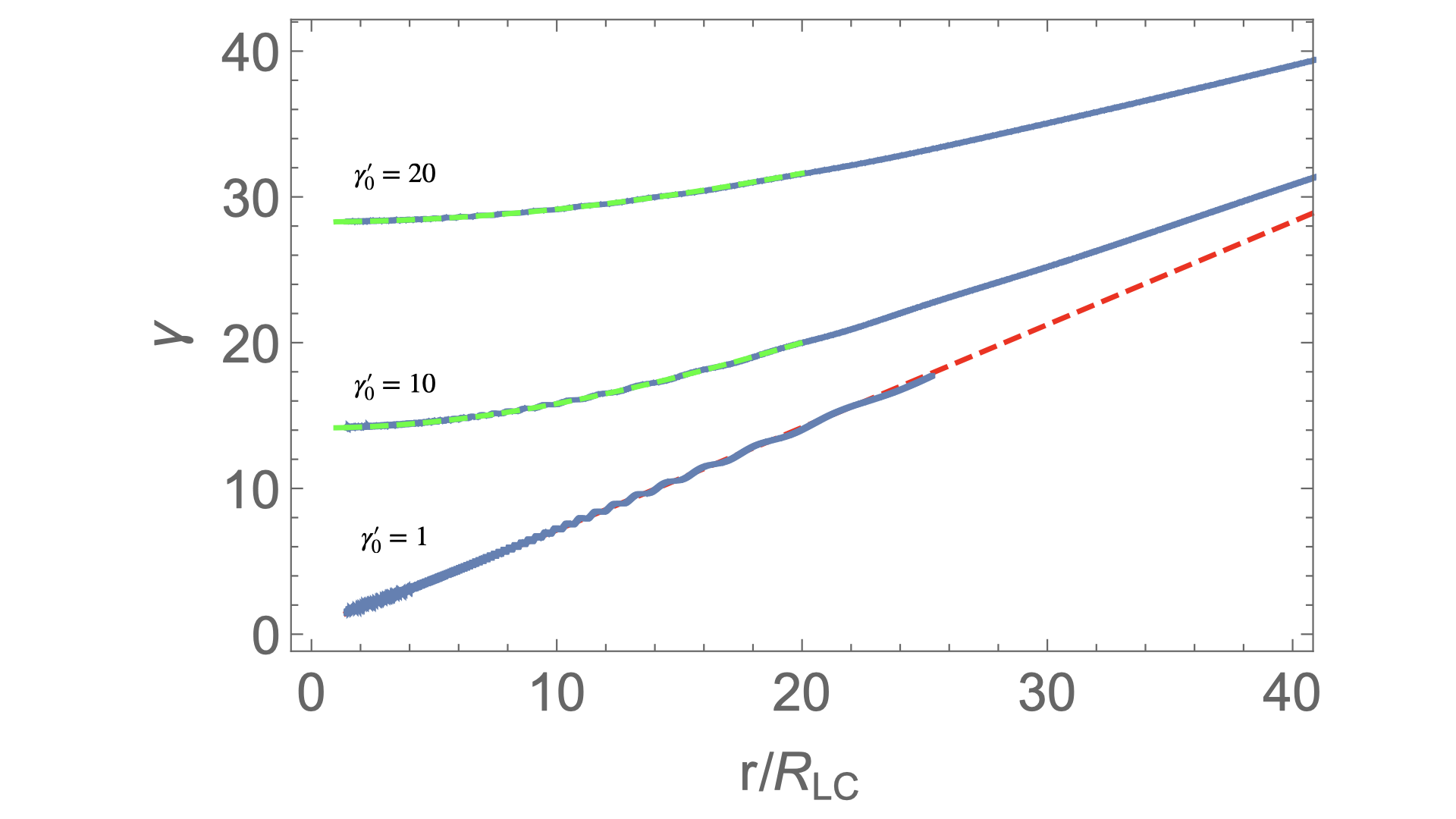}
    \caption{Direct integration  of particle trajectories for the \Lf\ of parallel motion  $\gamma_0'= 1, \, 10, \, 20$, $\theta = \pi/4$ (solid lines).  The $\gamma_0'=1$  reproduces Michel's solution $\gamma = \sqrt{1+r^2/2}$ (red dashed line). Green dashed lines: analytical solutions (\ref{gamma}).  In the simulations the \Bf\ at the \LC\ satisfies $\om_B = 10^ 4 \Omega$.}
\label{geom}
    \end{figure}

    \section{Jet in M87}

 \subsection{Particle dynamics in parabolic \Bf}
 \label{parabolicc}
 
 In our model collimation is completely separated from acceleration in the vicinity of the \LC. To demonstrate this, let's choose prescribed collimated flow along 
  parabolic flux surfaces
 \be
 r(1-\cos\theta) = {\rm const}
 \label{rrr}
 \ee
 In this case it is more convenient to work in cylindrical coordinates. The flux surface is then given by 
 \be
 z= \frac{\varpi^2- r^2_0}{2 r_0} 
   \ee
   ($\varpi$ is a cylindrical radial coordinate, $r_0$ is a parameter that marks a particular flux surface).
   
   In the metric tensor we first change to the rotating frame, $d\phi \to d\phi' - \Omega dt$, 
   add parabolic constraint $dz\to (\varpi/r_0) d \varpi$, and add a spiral step
   \ba &&
   d\phi' = \Omega dl
   \nn &&
   dl = \sqrt{ 1+ \varpi^2/r_0^2} d \varpi
   \ea
  At each location the velocity in $\{\varpi,\phi,z\}$ coordinates is 
  \be
  \left\{\beta _{\varpi },\Omega  \varpi  \left(1-\sqrt{1+\frac{\varpi ^2}{\varpi _0^2}}
   \beta _{\varpi }\right),\frac{\varpi  \beta _{\varpi }}{\varpi _0}\right\}
    \label{betavarpi}
   \ee
   
   Metric tensor in this case
   \ba &&
   G_{00} = 1- \Omega ^2 \varpi ^2
   \nn &&
   G_{0 \varpi} = 
   \Omega ^2 \varpi ^2 \sqrt{1+\frac{\varpi ^2}{\varpi _0^2}}
   \nn &&
   G_{\varpi \varpi} =
   \left(1+\frac{\varpi ^2}{\varpi _0^2}\right) \Omega ^2 \varpi ^2
   \ea

   Following our procedure we find
   \be
   \beta_\varpi =
   \frac{1-\left(2 \varpi-r_0\right) r_0 \Omega ^2}{\frac{\gamma _0}{\sqrt{\frac{\left(2
   \varpi -r_0\right) \left(\gamma _0^2+\left(2 \varpi-r_0\right) r_0 \Omega ^2-1\right)}{\left(2
   \varpi-r_0\right) r_0^2 \Omega ^2-2 \varpi}}}-\left(2 \varpi-r_0\right) r_0 \Omega ^2}
   \label{betavarpi1}
   \ee
  
\begin{figure}[!htb]
\includegraphics[width=0.99\linewidth]{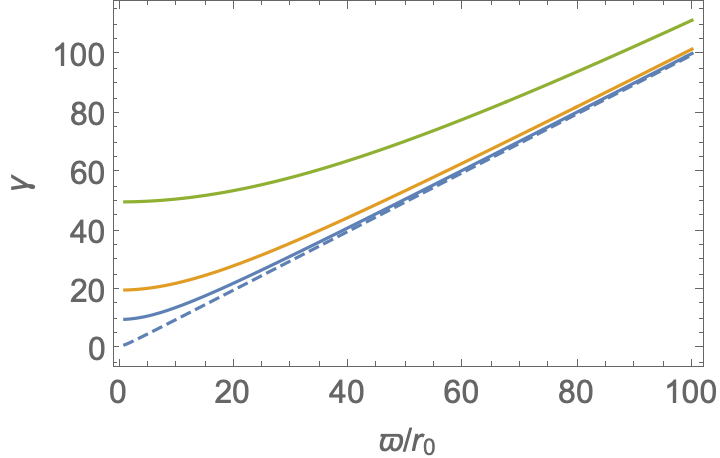}
    \caption{Lorentz factors of particles moving along  parabolic spiral for different $\gamma_0= 10,20,50$ (bottom to top curves). Dashed line is the \cite{1973ApJ...180L.133M} solution.}
\label{parabolic}
    \end{figure}

  The resulting acceleration differs little from the case of conical flux surfaces of the \cite{1973ApJ...180L.133M} solution, Fig. \ref{parabolic} 
    
\subsection{Emission maps}

As we demonstrated above, particles injected from the \BH\ \ms\ stream nearly along the magnetic flux surfaces - either conical for monopolar fields, parabolically, or more generally, along any given flux surface. The toroidal velocity remains small. 

For a  given shape of the magnetic flux surface (and some prescription for emissivity) we can then calculate the expected emission map. 
For monopolar \ms\ any emission by relativistic radially moving particles will be centered on the source, the \BH.

Even for curved flux surfaces, particles move with constant Lorentz factor almost purely along the flux surfaces, with minimal $v_\phi$.
For high injection {\Lf}s the emission will be dominated by the flow points when particle motion is aligned with the line of sight. 

For example, for parabolical flux surfaces \citep{1977MNRAS.179..433B}, also Eq. (\ref{rrr}),  neglecting GR contribution, the line of sight can be parallel to a given flux surface emanating from the \ms\
only if the lone of sight is $\theta_{ob} \leq \pi/4$. For smaller $\theta_{ob}$, for a given flux surface parameterized by $r_0$ the tangent point is 
\ba &&
\varpi_t = r_0 \cot \theta_{ob}
\nn &&
z_t= (\cot^2 \theta_{ob}-1) \frac{r_0}{2}
\ea

Choosing axis $x_s$ on the plane of the sky along the projection of the spin of the \BH\ on the plane of the sky, the tangent point projects to
\ba &&
x_s = \frac{r_0}{2 \sin \theta_{ob}}
\nn &&
y_s = 0
\label{parra}
\ea

\begin{figure}[!htb]
\includegraphics[width=0.99\linewidth]{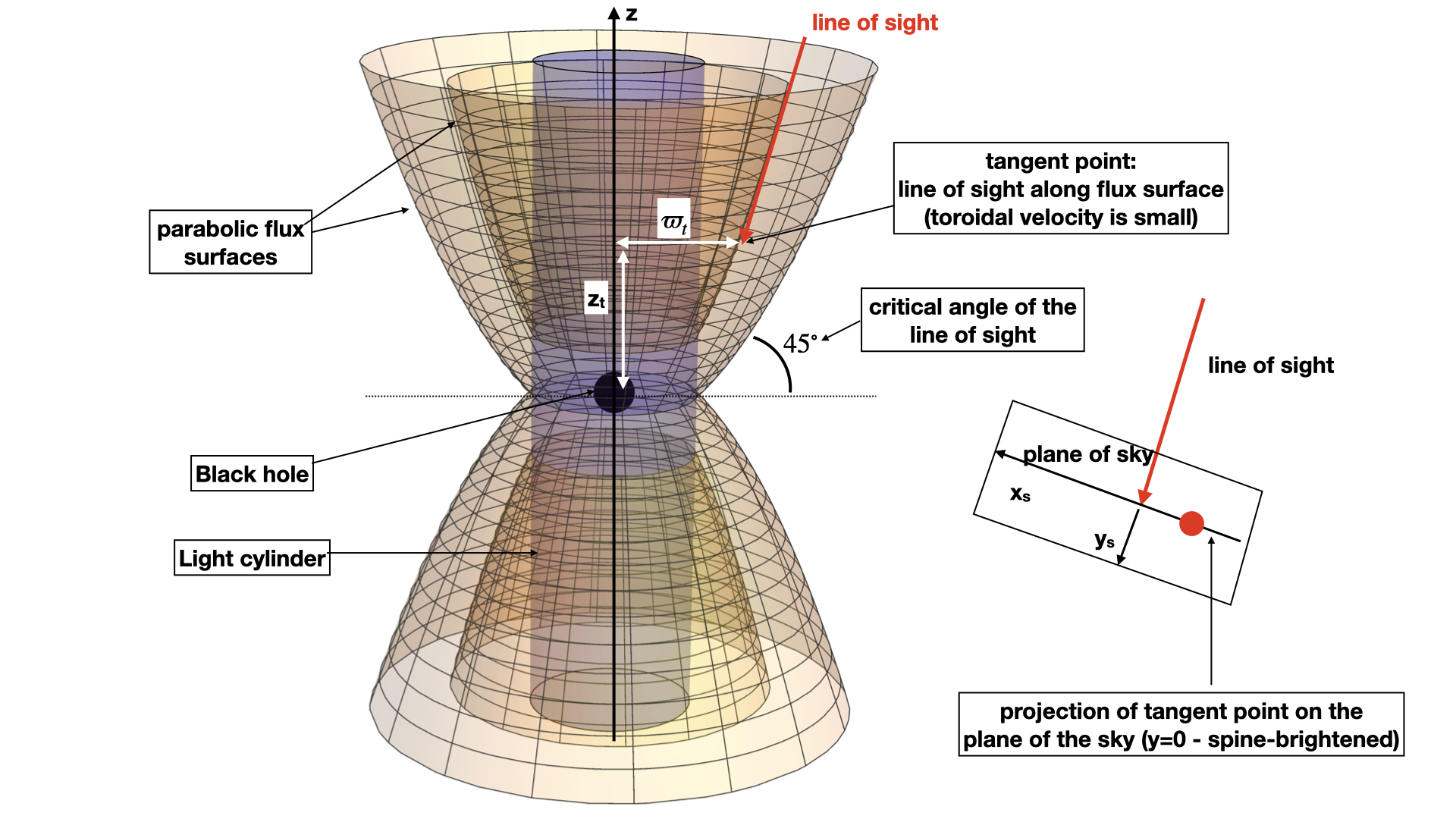}
    \caption{3D rendering of parabolic \ms. The vertical cylinder is the \LC. The insert on the right shows the sky image for a particular chosen line of sight. Due to high Doppler boosting the brightness peaks on the $y$ axis - the spine.}
\label{IC-BH002}
    \end{figure}

\begin{figure}[h!]
\includegraphics[width=0.85\linewidth]{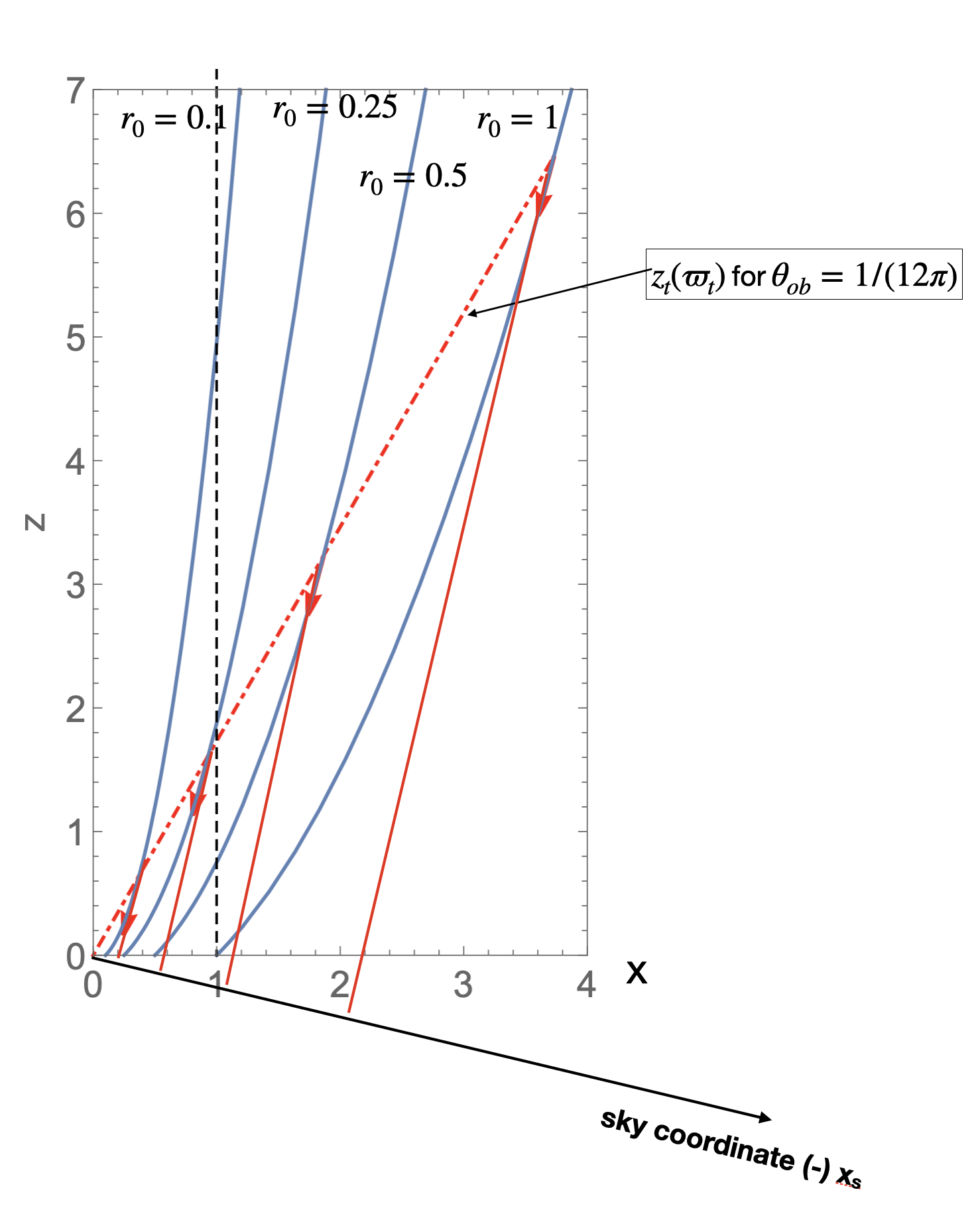}
    \caption{Parabolic \ms\ and the lines of sight, $x-z$ cut. Solid lines are magnetic flux surface for $r_0 =0.1, \, 0.25, \, 0.5, \, 1 $. Red arrows start at points where the local line of sight is along the flux surface (hence maximal Doppler factor). Red dashed line, Eq. (\ref{parra}) - location of all such points. In this example $\theta_{ob} =\pi/12$.
    Coordinates are normalized to \LC\ radius.}
\label{IC-BH003}
    \end{figure}

Since the observed pattern is dominated by Lorentz boost,
to calculate the images we employ the following procedure:
\begin{itemize}
\item Given the velocity (\ref{betavarpi}-\ref{betavarpi1}) we calculate local  Doppler factor $\delta$. In fact, since  toroidal velocity is small, one can use just the shape of the flux surface to find the local direction of the flow.  
\item We scale local density (somewhat arbitrary) as $1/r^2$, total spherical distance to the \BH.
\item local emissivity is parameterized as
\be
j \propto \frac{\delta^3}{r^2}
\ee
\item emissivity is integrated along the line of sight.
\end{itemize}

\begin{figure}[!htb]
\includegraphics[width=0.99\linewidth]{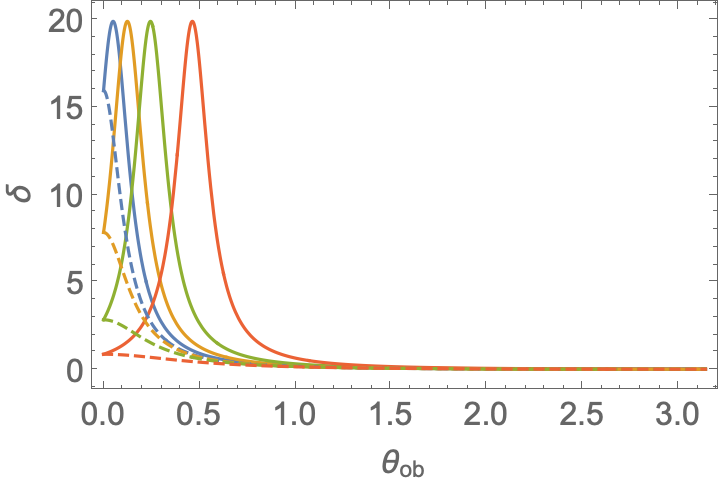}
    \caption{Doppler factor as a function of the observer angle $\theta_{ob}$  at $r=2$. Slices are   in the line-of-sight-\BH\ spin plane (solid lines) and   in the orthogonal plane (dashed lines). Peaks in $\delta$ correspond to angles when the line-of-sight is tangential to the flux surface. Here and for the images  below $\gamma_0 =10$.}
\label{deltaphitheta}
    \end{figure}
    
\begin{figure}[!htb]
\includegraphics[width=0.49\linewidth]{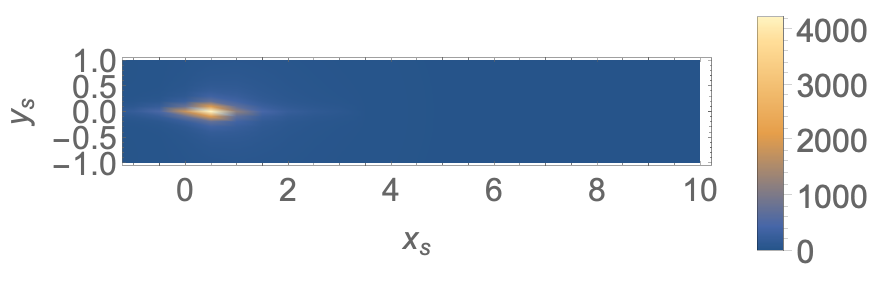}
\includegraphics[width=0.49\linewidth]{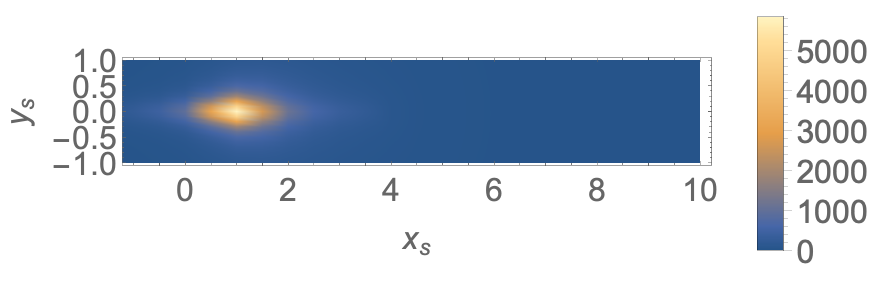}
\includegraphics[width=0.49\linewidth]{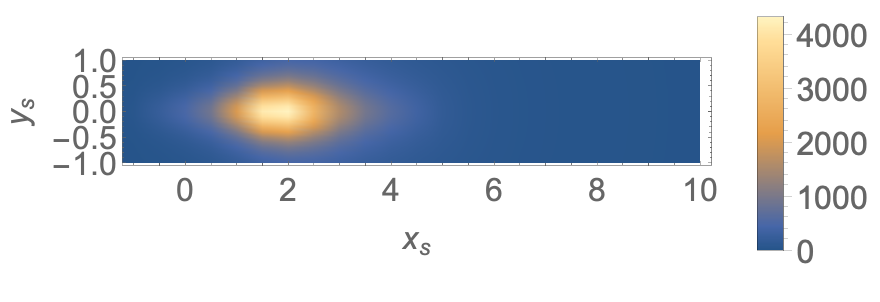}
\includegraphics[width=0.49\linewidth]{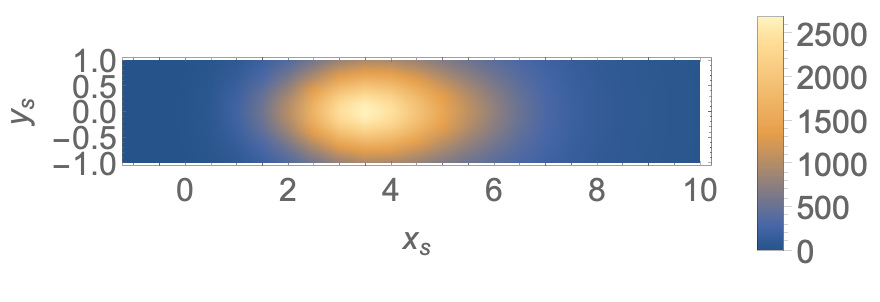}
    \caption{Image maps for $r_0 =0.1,\, 0.25, \, 0.5,\, 1$, $\theta_{ob} =\pi/12$. The axes correspond to projected distances  measured in terms of the \LC\ radius.}
\label{pi12}
    \end{figure}
    
\begin{figure}[!htb]
\includegraphics[width=0.49\linewidth]{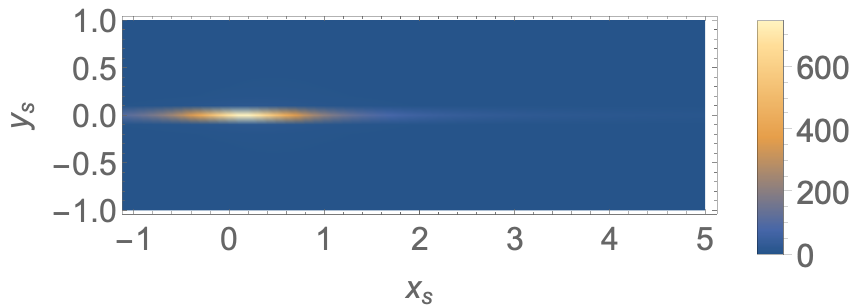}
\includegraphics[width=0.49\linewidth]{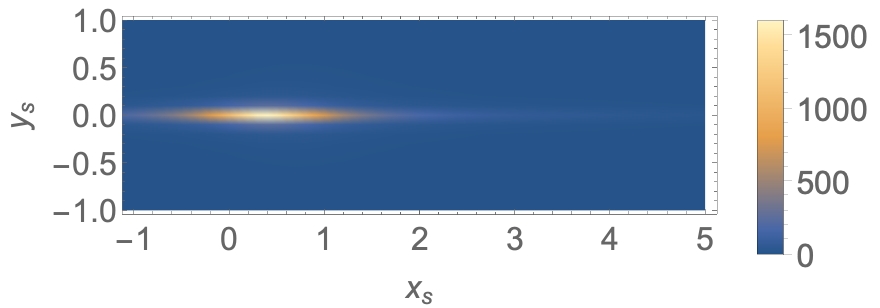}
\includegraphics[width=0.49\linewidth]{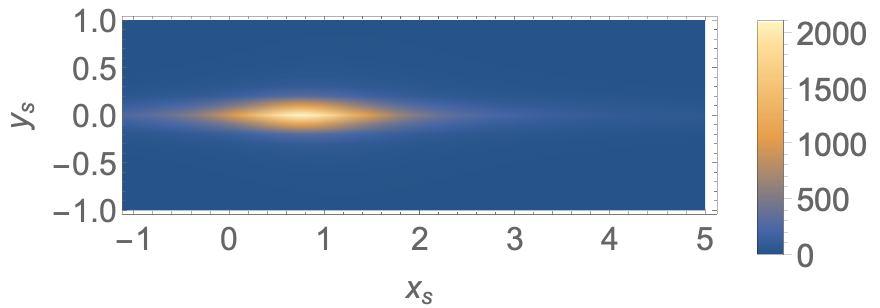}
\includegraphics[width=0.49\linewidth]{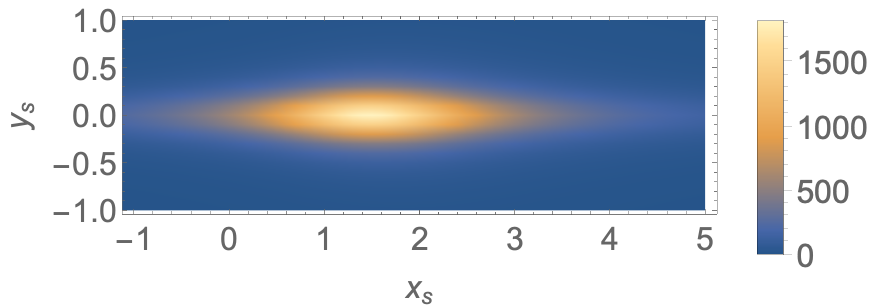}
    \caption{Image maps for $r_0 =0.1,\, 0.25, \, 0.5,\, 1$, $\theta_{ob} =\pi/6$.}
\label{pi6}
    \end{figure}

\begin{figure}[!htb]
\includegraphics[width=0.99\linewidth]{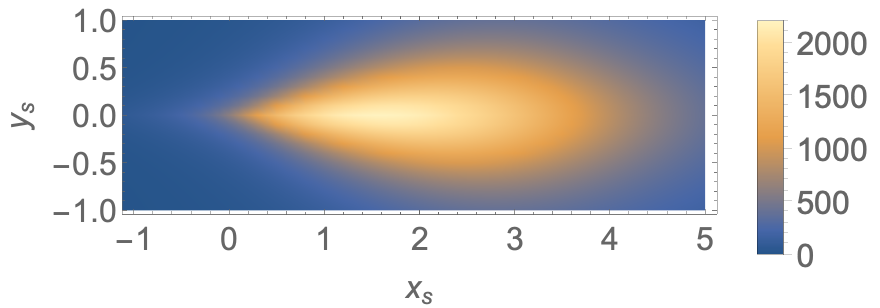}\\
\includegraphics[width=0.99\linewidth]{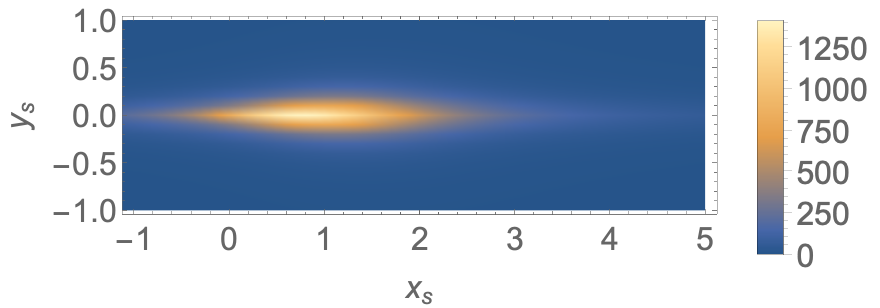}
    \caption{Image maps integrated over $0\leq r_0\leq 1$ for $\theta_{ob} =\pi/12$ (top panel) and $\theta_{ob} =\pi/6$ (bottom panel). Larger $r_0$ corresponds to flux surfaces not emanating from the \ms.}
\label{pi61}
    \end{figure}

In Fig. \ref{deltaphitheta}
    we plot slices of the Doppler factor in the two orthogonal planes. 
    Expected brightness maps are plotted in Figs. \ref{pi12}-\ref{pi61}.
All images, for any surface parameter $r_0$, are spine-brightened. Hence any combination will be spine-brightened as well.

\begin{figure}[!htb]
\includegraphics[width=0.99\linewidth]{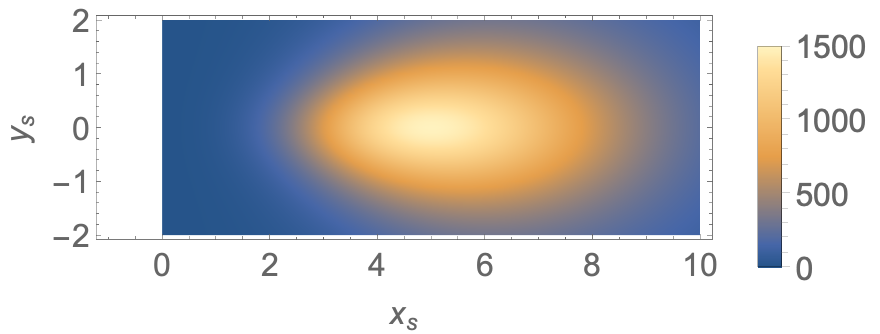}
    \caption{Emission pattern from the outside current sheet, $1\leq r_0\leq 2$ }
\label{outsided}
    \end{figure}

Finally, we conciliate emission from a possible current sheet outside of the \LC\ \citep{2021PhRvD.103b3014C}, by integrating the flux parameter $r_0$ from $1$ to $2$. The resulting structure is more  elongated, more extended sideways,  but is still spine-brightened, Fig. \ref{outsided}.

\subsection{Conclusion: morphology of M87 jet and the BZ mechanism}
Our results show a universal property of the resolved \cite{1977MNRAS.179..433B} flow with large parallel (to the local \Bf) momentum of emitting particles: all images are spine-brightened, {agreeing with what was} observed.
No special prescription for emissivity as function of the flux function parameter $r_0$ can change that: they are all  spine-brightened. The assumption of parabolical flux surfaces is, naturally, an analytic approximation, yet the  universality of the result - spine-brightened profile - ensures that it will be applicable to more general cases.

We then conclude that \cite{1977MNRAS.179..433B}
mechanism is {responsible} for the M87 jet, at least in its pure form. {\bf  By {\it BZ mechanism}  we understand generation of collimated relativistic jet. The relativistic  $e^\pm$ we describe do extract energy from the spin of the \BH.}

The \cite{1977MNRAS.179..433B} mechanism can still  be operational - and, \eg\ responsible for the bright core in the images of M87 in case of purely radial outflow at small $r$, but it does not drive the observed jet. Another possibility is that the sheath is slowed down by the interaction with the disk corona - while the core remains relativistic, with emission beamed away. This would correspond to emission coming only from small $r_0$, top rows in Figs. \ref{pi12} and \ref{pi6} - it is still spine-brightened, but shows  only in the small part of the image.

  In contrast disk-produced outflows \citep{1982MNRAS.199..883B} start non-relativistically. The extended structure observed in M87 is thus inconsistent with the \ms-produced jet, but is consistent with the disk-produce jets.
  
On larger scales, of the order of parsecs, in the grand spiral paradigm  of a jet  moving with \Lf\ of $\sim$ few \cite{2005MNRAS.360..869L},  the edge-brightened jet is actually expected on theoretical grounds \cite[\eg\ Fig. 1 of][]{2011MNRAS.415.2081C}. In addition, asymmetries across the jet  in intensity, polarization and spectral index maps are expected.
\subsection{Plasma dynamics and pair production in force-free and PIC  simulations}

Two approaches are commonly used to study relativistic winds and jets: force-free simulation  \citep[\eg]{2006ApJ...648L..51S} and PIC simulations \citep[\eg][]{2014ApJ...795L..22C,2015ApJ...801L..19P,2020PhRvL.124n5101C,2022arXiv220902121H,2022ApJ...924L..32R}. 

In ideal force-free simulation the velocity along the field is not defined in principle. (Inclusion of resistivity in the force-free approach requires some choice of parallel velocity \cite{2003MNRAS.346..540L,2007arXiv0710.1875G}.)
Thus the effects discussed here are completely missed in 
force-free simulation. 

In case of PICs the effects of initial large injection \Lf\ 
are missed by choice \citep[see though][]{2020PhRvL.124n5101C}.  To simulate pair production in PICs typically some kind of  prescription is employed: either a density floor, or a condition on the value of the local \Ef. Acceding to  a given prescription pairs are added {\it at rest}  \citep[except in][]{2020PhRvL.124n5101C}.  Instead, they should be added with large \Lfs. If the number of added pairs is comparable to the initial one, the resulting structure of the \ms\ will be drastically different.

Our results have implications for   PIC simulations  (that the newly born  $e^\pm$ pairs   should injected with  large  initial \Lf), and interpretation of the images of black holes by the Event Horizon Telescope.

This is a qualitatively different set-up from what the conventional  \BH\ acceleration models assume. Jets star fast, with \Lf\  $\gg 1$. At first they carry small power. Only later on the \Bf\ acceleration takes over. 

\section{Discussion}

We demonstrate that   outflows produced within \mss\ of  \NSs\ and  \BHs\ start relativistically right from within  the  \LC. To produce jets the overall \Bf\ just needs to collimate  the outflow - acceleration is already achieved by injection. For highly magnetized flows with $\sigma \gg 1$
the kinetic luminosity of the jet is $
L_k \sim L_P/\sigma \leq L_P $, when $L_P$ is the Poynting power,  but the jet is already relativistic. It can be further accelerated by the magnetic forces beyond $ r \Omega \sim \gamma_0$. 

Any signal produced by radially moving particles with $\gamma \gg 1$ will be centered on the  source. For parabolic flux surfaces  we calculated the expected images -  they are all spin-brightened, {in agreement} to observations of M87 jet. We then conclude that M87 jet {can} be produced with the \BH\ \ms\ by the BZ mechanism.

Finally, we point out that in PIC simulations pairs must be injected relativistically - this will change the overall structure of both pulsar and \BHs\ \mss.

\section*{Acknowledgments}
   This work had been supported by   
NASA grant  80NSSC18K064 and  NSF grants 1903332 and  1908590.
We would like to thank Nahum Arav, Maxim Barkov, Vasily Beskin, Ioannis Contopoulos, Hayk Hakobyan, Sergey Komissarov, Mikhlail Medvedev, Frank Rieger, Markek Sikora, Elena Nokhrina for comments,  and organizers of the IAU conference  "Black Hole Winds at all Scales" for hospitality.


\section{Data availability}
The data underlying this article will be shared on reasonable request to the corresponding author.



\bibliography{BibFiles}{}

\begin{thebibliography}{50}
\expandafter\ifx\csname natexlab\endcsname\relax\def\natexlab#1{#1}\fi

\bibitem[{{Bardeen} {et~al.}(1972){Bardeen}, {Press}, \& {Teukolsky}}]{Bardeen72}
{Bardeen}, J.~M., {Press}, W.~H., \& {Teukolsky}, S.~A. 1972, \apj, 178, 347

\bibitem[{{Barkov} \& {Komissarov}(2008)}]{2008MNRAS.385L..28B}
{Barkov}, M.~V., \& {Komissarov}, S.~S. 2008, MNRAS, 385, L28

\bibitem[{{Beskin}(2009)}]{BeskinBook}
{Beskin}, V.~S. 2009, {MHD Flows in Compact Astrophysical Objects: Accretion, Winds and Jets}

\bibitem[{{Beskin} {et~al.}(1992){Beskin}, {Istomin}, \& {Parev}}]{1992SvA....36..642B}
{Beskin}, V.~S., {Istomin}, Y.~N., \& {Parev}, V.~I. 1992, \sovast, 36, 642

\bibitem[{{Beskin} \& {Kuznetsova}(2000)}]{2000NCimB.115..795B}
{Beskin}, V.~S., \& {Kuznetsova}, I.~V. 2000, Nuovo Cimento B Serie, 115, 795

\bibitem[{{Bicknell} \& {Begelman}(1996)}]{1996ApJ...467..597B}
{Bicknell}, G.~V., \& {Begelman}, M.~C. 1996, \apj, 467, 597

\bibitem[{{Birdsall} \& {Langdon}(1991)}]{birdsall}
{Birdsall}, C.~K., \& {Langdon}, A.~B. 1991, {Plasma Physics via Computer Simulation}

\bibitem[{{Blandford} {et~al.}(2019){Blandford}, {Meier}, \& {Readhead}}]{2019ARA&A..57..467B}
{Blandford}, R., {Meier}, D., \& {Readhead}, A. 2019, \araa, 57, 467

\bibitem[{{Blandford} \& {K{\"o}nigl}(1979)}]{1979ApJ...232...34B}
{Blandford}, R.~D., \& {K{\"o}nigl}, A. 1979, \apj, 232, 34

\bibitem[{{Blandford} \& {Payne}(1982)}]{1982MNRAS.199..883B}
{Blandford}, R.~D., \& {Payne}, D.~G. 1982, \mnras, 199, 883

\bibitem[{{Blandford} \& {Znajek}(1977)}]{1977MNRAS.179..433B}
{Blandford}, R.~D., \& {Znajek}, R.~L. 1977, \mnras, 179, 433

\bibitem[{{Boris} \& {Roberts}(1969)}]{boris_69}
{Boris}, J.~P., \& {Roberts}, K.~V. 1969, Journal of Computational Physics, 4, 552

\bibitem[{{Camenzind}(1986)}]{1986A&A...162...32C}
{Camenzind}, M. 1986, \aap, 162, 32

\bibitem[{{Chen} \& {Beloborodov}(2014)}]{2014ApJ...795L..22C}
{Chen}, A.~Y., \& {Beloborodov}, A.~M. 2014, \apjl, 795, L22

\bibitem[{{Clausen-Brown} {et~al.}(2011){Clausen-Brown}, {Lyutikov}, \& {Kharb}}]{2011MNRAS.415.2081C}
{Clausen-Brown}, E., {Lyutikov}, M., \& {Kharb}, P. 2011, MNRAS, 415, 2081

\bibitem[{{Comisso} \& {Asenjo}(2021)}]{2021PhRvD.103b3014C}
{Comisso}, L., \& {Asenjo}, F.~A. 2021, \prd, 103, 023014

\bibitem[{{Contopoulos} {et~al.}(2020){Contopoulos}, {P{\'e}tri}, \& {Stefanou}}]{2020MNRAS.491.5579C}
{Contopoulos}, I., {P{\'e}tri}, J., \& {Stefanou}, P. 2020, \mnras, 491, 5579

\bibitem[{{Crinquand} {et~al.}(2020){Crinquand}, {Cerutti}, {Philippov}, {Parfrey}, \& {Dubus}}]{2020PhRvL.124n5101C}
{Crinquand}, B., {Cerutti}, B., {Philippov}, A., {Parfrey}, K., \& {Dubus}, G. 2020, \prl, 124, 145101

\bibitem[{{Goldreich} \& {Julian}(1970)}]{1970ApJ...160..971G}
{Goldreich}, P., \& {Julian}, W.~H. 1970, \apj, 160, 971

\bibitem[{{Grad}(1967)}]{Grad1967}
{Grad}, H. 1967, Physics of Fluids, 10, 137

\bibitem[{{Gralla} \& {Jacobson}(2014)}]{2014MNRAS.445.2500G}
{Gralla}, S.~E., \& {Jacobson}, T. 2014, \mnras, 445, 2500

\bibitem[{{Gruzinov}(2007)}]{2007arXiv0710.1875G}
{Gruzinov}, A. 2007, arXiv e-prints, arXiv:0710.1875

\bibitem[{{Hakobyan} {et~al.}(2022){Hakobyan}, {Philippov}, \& {Spitkovsky}}]{2022arXiv220902121H}
{Hakobyan}, H., {Philippov}, A., \& {Spitkovsky}, Anatoly, {\it et al}. 2022, arXiv e-prints, arXiv:2209.02121

\bibitem[{{Hirotani} \& {Okamoto}(1998)}]{1998ApJ...497..563H}
{Hirotani}, K., \& {Okamoto}, I. 1998, \apj, 497, 563

\bibitem[{{Kim} {et~al.}(2018){Kim}, {Krichbaum}, \& {Lu}}]{2018A&A...616A.188K}
{Kim}, J.~Y., {Krichbaum}, T.~P., \& {Lu}, R.~S., {\it et al}. 2018, \aap, 616, A188

\bibitem[{{Komissarov}(2004)}]{2004MNRAS.350..427K}
{Komissarov}, S.~S. 2004, \mnras, 350, 427

\bibitem[{{Komissarov} {et~al.}(2009){Komissarov}, {Vlahakis}, {K{\"o}nigl}, \& {Barkov}}]{2009MNRAS.394.1182K}
{Komissarov}, S.~S., {Vlahakis}, N., {K{\"o}nigl}, A., \& {Barkov}, M.~V. 2009, \mnras, 394, 1182

\bibitem[{{Krolik}(1999)}]{Krolik:1999}
{Krolik}, J.~H. 1999, {Active galactic nuclei : from the central black hole to the galactic environment}

\bibitem[{{Landau} \& {Lifshitz}(1975)}]{LLII}
{Landau}, L.~D., \& {Lifshitz}, E.~M. 1975, {The classical theory of fields}

\bibitem[{{Levinson}(2000)}]{2000PhRvL..85..912L}
{Levinson}, A. 2000, \prl, 85, 912

\bibitem[{{Levinson} \& {Rieger}(2011)}]{2011ApJ...730..123L}
{Levinson}, A., \& {Rieger}, F. 2011, \apj, 730, 123

\bibitem[{Lu {et~al.}(2023)Lu, Asada, \& Krichbaum}]{lu2023ring}
Lu, R.-S., Asada, K., \& Krichbaum, Thomas~P, {\it et al}. 2023, Nature, 616, 686

\bibitem[{{Lyutikov}(2003)}]{2003MNRAS.346..540L}
{Lyutikov}, M. 2003, \mnras, 346, 540

\bibitem[{{Lyutikov}(2009)}]{2009MNRAS.396.1545L}
---. 2009, \mnras, 396, 1545

\bibitem[{{Lyutikov}(2022)}]{2022ApJ...933L...6L}
---. 2022, \apjl, 933, L6

\bibitem[{{Lyutikov} {et~al.}(2005){Lyutikov}, {Pariev}, \& {Gabuzda}}]{2005MNRAS.360..869L}
{Lyutikov}, M., {Pariev}, V.~I., \& {Gabuzda}, D.~C. 2005, MNRAS, 360, 869

\bibitem[{{McKinney}(2006)}]{2006MNRAS.368.1561M}
{McKinney}, J.~C. 2006, \mnras, 368, 1561

\bibitem[{{Michel}(1969)}]{1969ApJ...158..727M}
{Michel}, F.~C. 1969, \apj, 158, 727

\bibitem[{{Michel}(1973)}]{1973ApJ...180L.133M}
---. 1973, \apjl, 180, L133

\bibitem[{{Misner} {et~al.}(1973){Misner}, {Thorne}, \& {Wheeler}}]{MTW}
{Misner}, C.~W., {Thorne}, K.~S., \& {Wheeler}, J.~A. 1973, {Gravitation} (San Francisco: W.H.~Freeman and Co., 1973)

\bibitem[{{Nakamura} \& {Asada}(2013)}]{2013ApJ...775..118N}
{Nakamura}, M., \& {Asada}, K. 2013, \apj, 775, 118

\bibitem[{{Nokhrina} \& {Beskin}(2017)}]{2017MNRAS.469.3840N}
{Nokhrina}, E.~E., \& {Beskin}, V.~S. 2017, \mnras, 469, 3840

\bibitem[{{Philippov} {et~al.}(2015){Philippov}, {Spitkovsky}, \& {Cerutti}}]{2015ApJ...801L..19P}
{Philippov}, A.~A., {Spitkovsky}, A., \& {Cerutti}, B. 2015, \apjl, 801, L19

\bibitem[{{Prokofev} {et~al.}(2018){Prokofev}, {Arzamasskiy}, \& {Beskin}}]{2018MNRAS.474.1526P}
{Prokofev}, V.~V., {Arzamasskiy}, L.~I., \& {Beskin}, V.~S. 2018, \mnras, 474, 1526

\bibitem[{{Ptitsyna} \& {Neronov}(2016)}]{2016A&A...593A...8P}
{Ptitsyna}, K., \& {Neronov}, A. 2016, \aap, 593, A8

\bibitem[{{Ripperda} {et~al.}(2022){Ripperda}, {Liska}, \& {Chatterjee}}]{2022ApJ...924L..32R}
{Ripperda}, B., {Liska}, M., \& {Chatterjee}, K., {\it et al}. 2022, \apjl, 924, L32

\bibitem[{{Scharlemann} \& {Wagoner}(1973)}]{1973ApJ...182..951S}
{Scharlemann}, E.~T., \& {Wagoner}, R.~V. 1973, \apj, 182, 951

\bibitem[{{Shafranov}(1966)}]{Shafranov1966}
{Shafranov}, V.~D. 1966, Reviews of Plasma Physics, 2, 103

\bibitem[{{Spitkovsky}(2006)}]{2006ApJ...648L..51S}
{Spitkovsky}, A. 2006, \apjl, 648, L51

\bibitem[{{Tomimatsu}(1994)}]{1994PASJ...46..123T}
{Tomimatsu}, A. 1994, \pasj, 46, 123

\end{thebibliography}
\bibliographystyle{apj}

\appendix
\section{Constrained motion  in flat space}
\label{Lagrangianwithconstraint}

Here we aim to re-derive the results for flat metric using Lagrangian and Hamiltonian approaches for constrained motion in flat space.

Consider relativistic particle moving along rotating spiral in cylindrical coordinates  in flat space. One possible choice of  relativistic Lagrangian is \citep{LLII}
\be
{\cal L} = - \sqrt{1-\beta_\phi^2 - \beta_r^2}
\label{LL1}
\ee
(see also Appendix \ref{TheLagrangianApproachAppE} for an alternative choice).

\begin{figure}[!htb]
\includegraphics[width=0.49\linewidth]{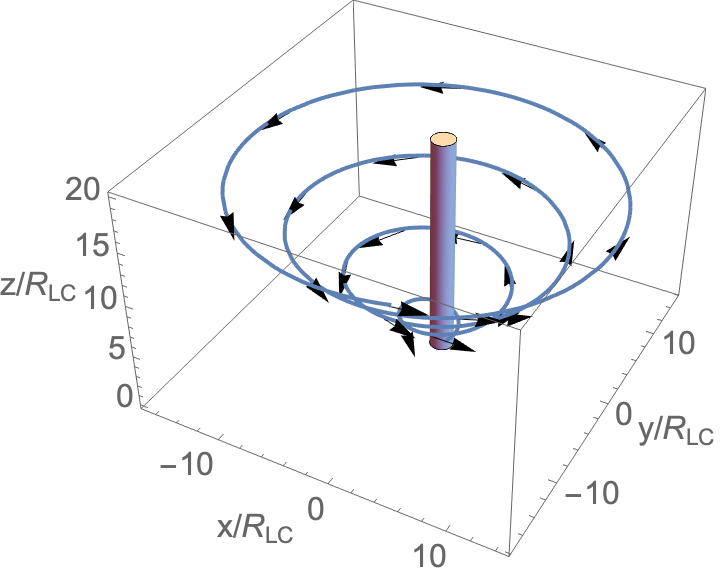}
\includegraphics[width=0.49\linewidth]{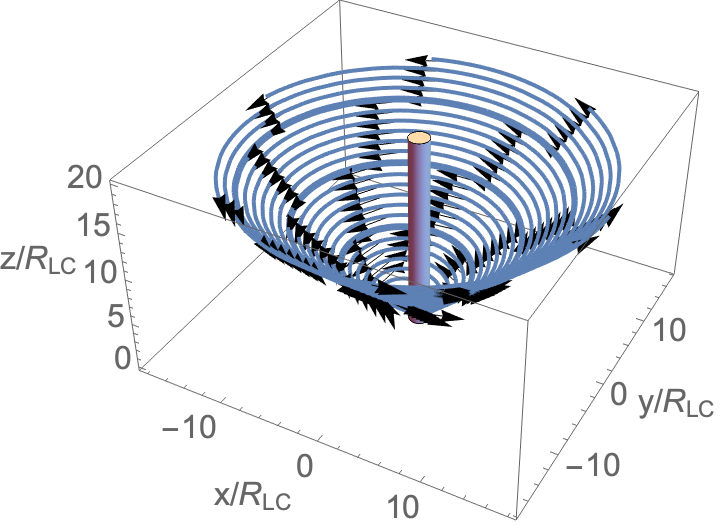}
    \caption{ Field lines for   Michel's solution.}
\label{radial22}
    \end{figure}

 The  field lines of the Michel's solution are given as a parametric curve in $x-y-z$ coordinates (at any  moment $t$)
 \ba &&
{\cal C}: \{r \sin (\theta ) \cos (\phi ),r \sin (\theta ) \sin (\phi ),r \cos (\theta )\}
 \nn && 
 \phi =  ( t- r  ) \Omega
 \label{CC}
 \ea
 where spherical $r$ is a parameter along the curve, Fig. \ref{radial22}.
 
Total velocity (differentiating  (\ref{CC}) with respect to time)
   \be\label{SpeedFlatArch}
   {\bf v} = \beta_r {\bf e}_r + \sin \theta r \Omega(1-\beta_r ) e_\phi
   \ee
   The Lagrangian then becomes
   \be
    {\cal L} = - 
    \sqrt{1 -\beta_r^2 - \sin^2 \theta r^2 \Omega^2 (1-\beta_r)^2}
    \label{LL}
    \ee
    ($\theta=\pi/2$ is assumed  below for simplicity).

 Lagrange equation for coordinate $r$ with $\beta_r = \partial_t r$
\be
\partial _ r {\cal L} = \partial_t (\partial _ {\beta_r}  {\cal L}  )
\ee
gives
\be 
\partial_t \beta_r = r \Omega^2 (1-\beta_r) ^2( 1 - r ^2 \Omega^2 +\beta_r (2 + r^2 \Omega^2))
\label{acc}
\ee


Solution satisfying $r(t=0) = 1/\Omega$  is 
\be
t=r-\frac{1}{\Omega }+ \frac{\log \left(\frac{\left(\gamma _0^2-1\right) (r \Omega -1) \left(\sqrt{\gamma
   _0^2+r^2 \Omega ^2-1}+\gamma _0 r \Omega \right)}{\gamma _0^2 (r \Omega +1)
   \left(\gamma _0 r \Omega -\sqrt{\gamma _0^2+r^2 \Omega ^2-1}\right)}\right)}{2 \Omega
   }
   \label{trajjj} 
   \ee
where the integration constant $\gamma_0$  was chosen to match (\ref{brflat}).
Differentiating (\ref{trajjj}) with respect to time we recover (\ref{brflat}).

As a check, let us repeat the above derivation using Hamiltonian approach. 
Given Lagrangian (\ref{LL})
we can find  canonical momentum $P_r$ and Hamiltonian

\ba &&
P_r= \partial _{\beta_r} {\cal L}= \frac{\beta_r-r^2 \Omega ^2 \left(1-\beta _r\right)}{\sqrt{1 -\beta_r^2 - r^2 \Omega^2 (1-\beta_r)^2}}
\nn && 
{\cal H} ={\beta_r} P_r -  {\cal L} =
\frac{r^2 \Omega ^2 P_r+\sqrt{P_r^2+r^2 \Omega ^2+1}}{r^2 \Omega ^2+1} 
\nn && 
\ea

Canonical equations are then
\ba &&
\beta_r = \partial_t r = \partial_{P_r} {\cal H} =
\frac{r^2 \Omega ^2}{1+ r^2 \Omega ^2} +
\frac{P_r}{(1+ r^2 \Omega ^2) \sqrt{1+P_r^2 + ( r  \Omega)^2 } }
\nn &&
\partial_t P_r = -\partial_r {\cal H} =
\frac{r \Omega ^2 \left(-2 P_r \sqrt{P_r^2+r^2 \Omega ^2+1}+2 P_r^2+r^2 \Omega
   ^2+1\right)}{\left(r^2 \Omega ^2+1\right)^2 \sqrt{P_r^2+r^2 \Omega ^2+1}}
   \label{betarr}
\ea

Equating ${\cal H} = \gamma_0$
we find
\be 
P_r = \frac{\gamma _0 r^2 \Omega ^2+\sqrt{\gamma _0^2+r^2
   \Omega ^2-1}}{1-r^2 \Omega ^2} 
   \label{Prr}
\ee
Using (\ref{Prr}) in expression for velocity (\ref{betarr})  we recover (\ref{brflat}).

The Lagrangian approach  requires integration of the equations of motion and allows one to  find $r(t)$, while the more simple  Hamilton-Jacobi approach, which involves only algebraic relations, gives only $\beta_r(r)$. This is sufficient for our application.

\section{Another Lagrangian approach in  \Sc\ metric}
\label{TheLagrangianApproachAppE}

Let us next extend the results of Appendix \ref{Lagrangianwithconstraint} to \Sc\ metric using a somewhat different approach. We are  interested in getting the radial speed of the particle along the string (i.e., on the (1+1)-D manifold). The alternative choice of Lagrangian is
\begin{equation}
    \mathcal{L}(\dot{t},\dot{r}, r) = \frac{1}{2}\left[-G_{00}\, \dot{t}^2 + G_{rr}\, \dot{r}^2 + 2G_{0r}\,\dot{t}\dot{r}\right], \hspace{2cm} \dot{x} = \frac{dx}{d\tau}
\end{equation}
From the symmetry of the Lagrangian in the $t$ coordinate, we use the Euler-Lagrange equation to get
\begin{equation}\label{tDOT}
    \dot{t} = \frac{G_{0r}\, \dot{r} + E}{G_{00}\,},\hspace{2cm} G_{00}\, \neq 0
\end{equation} where $E$ is a constant of motion. It is generally different from $\gamma_0$ - different choices of the Lagrangian result in different integration constants; the exact relations between different constants are typically complicated.

For timelike trajectories,
\begin{equation}
    \mathcal{L}(\dot{r}, r) = \frac{1}{2}\left[\left(G_{rr}\, + \frac{G_{0r}^2}{G_{00}}\right)\dot{r}^2 - \frac{E^2}{G_{00}\,}\right] = -\frac{1}{2}
\end{equation}
The speed of the coordinate $r$ with respect to the proper time (the time measured by the particle that is constrained to move along the string) is
\begin{equation}\label{VelocitySquared2}
    \dot{r} = \pm \sqrt{\frac{E^2 - G_{00}\,}{ G_{0r}^2+G_{00}\,G_{rr}\, }}
\end{equation}
The sign here determines the direction of motion: either towards or away from the black hole. 
The speed of the coordinate $r$ with respect to the coordinate $t$ is
\begin{equation}\label{VelocitySquared2_t_tDot}
   \beta_r = \frac{dr}{dt} = \frac{\dot{r}}{\dot{t}} \:\overset{(\ref{tDOT})}{=}\:G_{00}\, \left(G_{0r}\, + \frac{E}{\dot{r}}\right)^{-1} = G_{00}\, \left(G_{0r}\, \pm E\sqrt{\frac{G_{0r}^2+G_{00}\,G_{rr}\, }{E^2-G_{00}\,}}\right)^{-1}
\end{equation}
This recovers equation (\ref{MAIN}), obtained by the Hamilton-Jacobi approach. 

Consider rotating \ms\ in  \Sc\ metric. 
We use the following convention
\begin{eqnarray}
G_{00}=\alpha^2-r^2\Omega^2, \hspace{0.5cm} G_{rr}=\frac{1}{\alpha^2}+\frac{r^2\Omega^2}{\alpha^4}, \hspace{0.5cm} G_{0r} = -\frac{r^2\Omega^2}{\alpha^2}
\end{eqnarray}
The Lagrangian is then
\ba \label{SCHWARZSCHILDLAGRANGIANAPPENDIXB} && 
{\cal L} =
 \frac{1}{2}\left(-\alpha^2\left(1-\frac{r^2\Omega^2}{\alpha^2}\right)\, \dot{t}^2 + \frac{1}{\alpha^2}\left(1+\frac{r^2\Omega^2}{\alpha^2}\right)\, \dot{r}^2 - 2\frac{r^2\Omega^2}{\alpha^2}\,\dot{t}\dot{r}\right)
\ea
Between the two light cylinders, the constant of motion
\be
E =\left(\alpha ^2-r^2 \Omega ^2\right)\dot{t} + \frac{r^2\Omega ^2}{\alpha ^2}\dot{r}
\label{gamma00}
\ee
We then get $\dot{r}$ by substituting in (\ref{VelocitySquared2}) or by solving the quadratic equation (\ref{SCHWARZSCHILDLAGRANGIANAPPENDIXB}) using $G_{00}G_{rr}+G_{0r}^2=1$ (in our case):
\begin{equation}
    \dot{r} = 
    \frac{-G_{0r}\dot{t}\pm\sqrt{\dot{t}^2-G_{rr}}}{G_{rr}}
\end{equation}
Eliminating $\dot{t}$ with (\ref{gamma00}),
\be
\dot{t}=\frac{\alpha ^2 E-r^2 \Omega ^2\dot{r}}{\alpha ^2 \left(\alpha ^2-r^2 \Omega ^2\right)}
\ee
we find Lagrangian and Hamiltonian
\ba &&
{\cal L} = \frac{\dot{r}^2-E^2}{2 \left(\alpha ^2-r^2 \Omega ^2\right)}
\nn &&
{\cal H} = \dot{r} \partial _{\dot{r}} {\cal L } - {\cal L } = \frac{\dot{r}^2+E^2}{2 \left(\alpha ^2-r^2 \Omega ^2\right)}
\ea

We used ($\sqrt{2}$ times) the proper time to parameterize the Lagrangian in (\ref{SCHWARZSCHILDLAGRANGIANAPPENDIXB}); thus, the Lagrangian has a constant value of $-1/2$ with this parameterization for timelike geodesics. Therefore, $\dot{r}$ is
\be
\dot{r}=\pm\sqrt{r^2 \Omega ^2+E^2-\alpha ^2},
\ee
consistent with (\ref{d2tau}).

The radial velocity measured by the coordinate time of the observer is then
\be
\beta_r = \frac{\dot{r}}{\dot{t}} =
\frac{\alpha ^2 \left(\alpha ^2-r^2 \Omega ^2\right)}{\frac{\alpha ^2 E}{\sqrt{r^2 \Omega ^2+E^2-\alpha ^2}}-r^2 \Omega ^2}
   \ee
   
The geodesic equation takes a nice form 
\be
\Ddot{r} = \frac{1}{2}\frac{d}{dr}\left(\dot{r}^2\right)=r \Omega^2 - \frac{M}{r^2},
\ee
(derivatives are with respect to proper time) which unites the cases of radial motion in \Sc\ geometry \cite[\eg][]{MTW} and (\ref{d2tau}).

We also note that one can get the same differential equation (for the case between the two light cylinders) by solving the classical mechanical problem of a particle with mass $m$ constrained to move along a straight wire rotating with a constant angular speed $\Omega$ in the flat 2-D plane, with the inclusion of a source of mass $M$ at the center. The Lagrangian is ($\dot{r}=dr/dt$)
\begin{equation}
\mathcal{L}=\frac{1}{2}m\left(\dot{r}^2+r^2\Omega^2\right)-U(r),\hspace{1cm} U(r)=-m\frac{MG}{r}
\end{equation}

\section{Limitations on angular velocity of field lines}
\label{limits}
\begin{figure}[h!]
\includegraphics[width=1\linewidth]{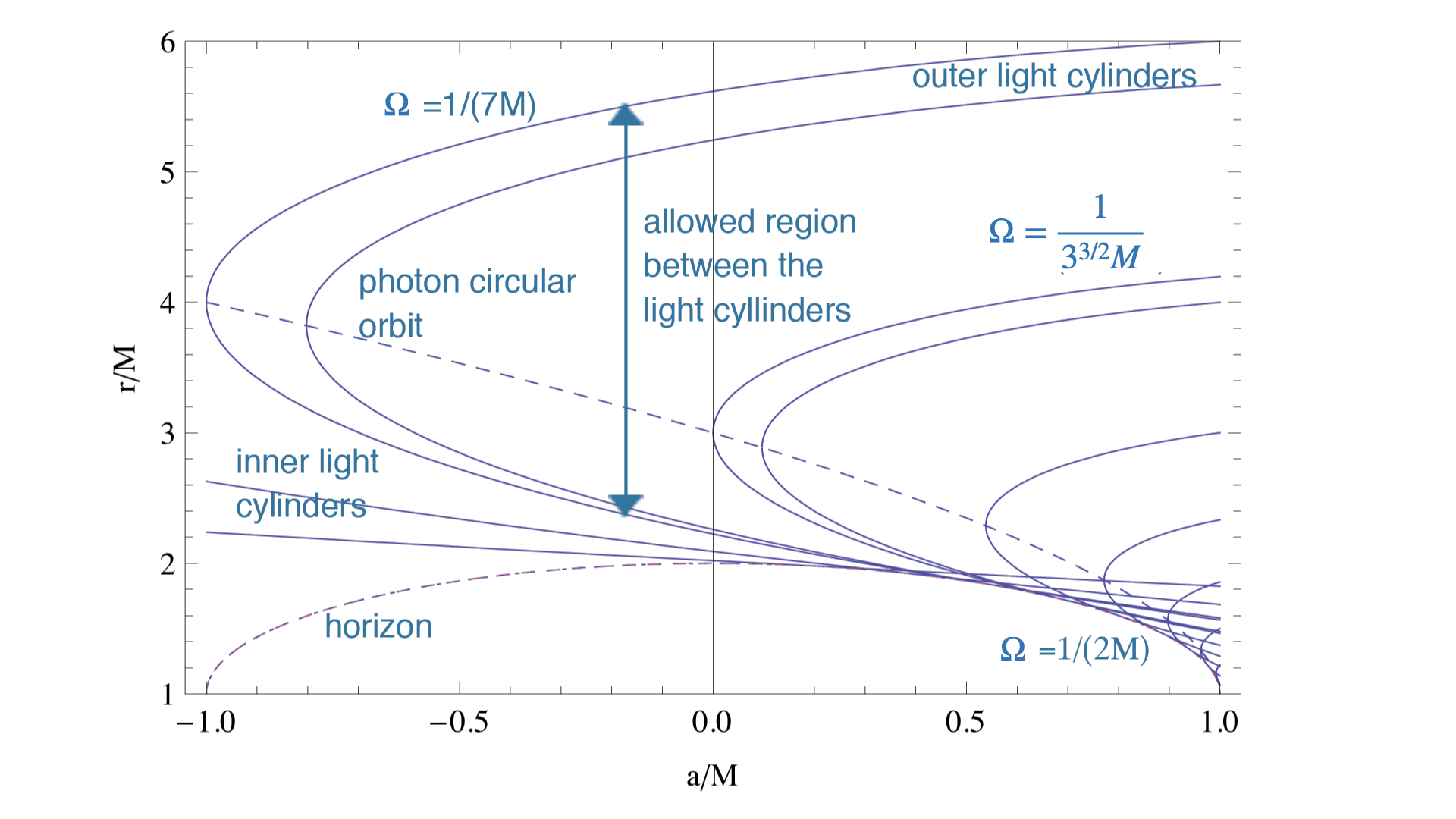}
\caption{The locations of {\LC}s for different $\om$ as function of $a$.  The light cylinders always lie outside the horizon (dotted curve). For a given $\om$, the inner and outer {\LC}s merge at  the photon circular orbit (dashed line).
 The physically meaningful region lies in between the  {\LC}s,   to the right of the {\LC}s curve. For $\om \rightarrow 0$, the inner \LC\  coincides with the ergosphere $r=2M$.  As $\om$ increases, the outer \LC\ moves to smaller $r$; the radial location of the  inner \LC\  is a complicated function of $\om$.  For $\om > 1/(7M)$ there is a region for sufficiently small $a$, for which the transformation to the rotating frame is unphysical.
For high $\om \rightarrow 1/(2M)$ and high $a > M/\sqrt{2}$, the inner \LC\ moves inside the ergosphere of a Kerr \BH.  
 For $\om > 1/(2M)$,   the transformation to rotating frame is  unphysical everywhere. Adapted from  \protect\citep{2009MNRAS.396.1545L}}
  \label{r(t)}
\end{figure}

 Transformation to the rotating Kerr metric has limitations \citep{2009MNRAS.396.1545L}:  it is physical only  for angular velocities smaller than $\om_{ph}$,  angular velocity of a photon orbit, defined by the conditions of circular rotation with the speed of light $g_{00}=0, \, \partial_r g_{00}=0$.  This gives $-4 a^2 M + r (-3 M + r)^2=0$ \citep{Bardeen72}. 
For a given $a$ and $r$, the transformation to the rotating Kerr metric   becomes meaningless for $\om$ higher than the angular velocity of a photon circular orbit, 
\be
\om_{\rm ph } = { 1 \over \mp | a| + 6 M \cos({1\over 3} {\rm arccos} (\mp |a|/M)}
\label{k2}
\ee
The upper sign corresponds to prograde rotation. Particular values are $a=0,\, r_{\rm ph}=3M,\, \om _{\rm ph}=1/(3\sqrt{3} M)$ for \Sc\ \BH, 
$a=M,\, r_{\rm ph}=M,\,\om_{\rm ph}=1/(2M)$ for prograde and $a=-M,\, r_{\rm ph}=4M,\,\om_{\rm ph}=1/(7M)$ for retrograde photon orbits.
For $1/(7M) < \om < 1/(2M) $ this requires  sufficiently high $a$, satisfying  $\om < 1/\left( a +2 \sqrt{2} M \left(1+\cos \left( {2\over 3}  {\rm arccos} (-{a\over M} ) \right)\right)\right)^{3/2}$ (for $a=0$ this requires $\om  < 1/(3 \sqrt{3})$). For $\om \rightarrow 1/ (2 M)$ both {\LC}s merge on the \BH\ horizon at  $a=M=r$.  For higher $\om $, the  transformation to the  rotating frame becomes  meaningless everywhere (Fig. \ref{r(t)}). 

\newpage
\section{Kerr Black Hole Analysis}\label{KERRBLACKHOLEAPPENDIXAHMAD}
\label{TheInvariantLorentzFactorAppendixE}
In this Appendix, we calculate the velocity components relative to an observer at infinity and show that the radial component dominates in a few light cylinder radii; we also obtain the Lorentz factor of the speed relative to both fixed observers in spacetime and co-rotating observers with the field lines.

We consider a (2+1)-D manifold of the Kerr black hole with a constant $\theta$. Then, we move to a (1+1)-D sub-manifold that is rotating with a constant angular speed $\Omega$ according to
\begin{equation}
    d\phi = \Phi'(r)\, dr + \Omega \, dt \hspace{2cm} \Phi'(r) = \frac{d\Phi(r)}{dr}
\end{equation}
This means that a particle in this sub-manifold is constrained to move along a wire with a shape function $\Phi(r)$. {Using the formula in (\ref{spiral}), we define the wire function as follows:} 
\begin{equation}
     {\Phi'(r) = \omega_{sp} =-\sqrt{\frac{g_{rr}}{g_{00}}}\left(\frac{g_{0\phi}}{g_{\phi\phi}}+\Omega_1(r)\right)}
\end{equation}
where $\Omega(r)$ is a function of the radial coordinate as a generalization of the Archimedean spiral shape (e.g., $\Omega_1(r)=\Omega$ is the Archimedean spiral case). We note from (\ref{PhysicallyRealizableStep}) that this function has some constraints to have a physically realizable spiral step (like in the Schwarzshild metric, $\Omega_1(r) \geq \Omega$).
Using the notation in (\ref{GG}), the metric of is
\begin{eqnarray}\label{KerrMetricGenelarized2}
&&ds^2 = -G_{00}\, dt^2 + G_{rr}\, dr^2 + 2 G_{0r}\, dt dr
\mbox{} \nonumber \\ \mbox{} &&
G_{00}\, = g_{00}\, -(\Omega g_{\phi \phi}\, + 2 g_{0 \phi}\,)\Omega \mbox{} \nonumber \\ \mbox{} && 
G_{rr}\, = g_{rr}\, +\Phi'(r)^{\,2} \,g_{\phi \phi}\,
\mbox{} \nonumber \\ \mbox{} && 
G_{0r}\, = \left(\Omega g_{\phi \phi}\, + g_{0 \phi}\,\right)\Phi'(r)
\end{eqnarray}

Note that we have constraints on the values of the mass-rotation parameter $\eta = M\Omega$ and the Kerr parameter $a$:
\begin{eqnarray}\label{SpecificCubicEquationConstraint}
&&
\eta \sqrt{\frac{(1-a\eta)}{(1+a\eta)^3}}< \frac{1}{3\sqrt{3}} 
\end{eqnarray}
{In the Kerr case, we have (by definition) the following coordinate basis vectors
\begin{equation}
    (\boldsymbol{e}_0)^\mu=\left(\begin{array}{cc}
        1\\
        0\\
        0
    \end{array}\right) \hspace{2cm} (\boldsymbol{e}_r)^\mu=\left(\begin{array}{cc}
        0\\
        1\\
        0
    \end{array}\right) \hspace{2cm} (\boldsymbol{e}_\phi)^\mu=\left(\begin{array}{cc}
        0\\
        0\\
        1
    \end{array}\right)
\end{equation}
Of course, these basis vectors satisfy $\boldsymbol{e}_\mu \cdot \boldsymbol{e}_\nu = g_{\mu\nu}$. The lengths are
\begin{equation}
    |\boldsymbol{e}_\mu| = \sqrt{g_{\mu\mu}} \hspace{1cm}\Rightarrow\hspace{1cm} |\boldsymbol{e}_0| = \sqrt{-g_{00}} \hspace{1cm} |\boldsymbol{e}_r| = \sqrt{g_{rr}} \hspace{1cm} |\boldsymbol{e}_\phi| = \sqrt{g_{\phi\phi}} 
\end{equation}}
{To get the velocity components, we need to take the total (intrinsic) derivative of the normalized position vector components $\hat{V}^\mu$ (note $v^\mu := dx^\mu/dt$)
\begin{equation}
    \frac{D}{Dt}\hat{V}^\mu = |\boldsymbol{e}{_\mu}{}| \left[\frac{d}{dt}\left(\frac{\hat{V}^\mu}{|\boldsymbol{e}{_\mu}{}|}\right) + \left(\Gamma^\mu_{\alpha\beta}v^\beta\right)\left(\frac{\hat{V}^\alpha}{|\boldsymbol{e}{_\alpha}{}|}\right)\right]
\end{equation}
We obtain the velocity components as follows
\begin{enumerate}
    \item The radial component $\beta_r$
    \begin{equation}
    \beta_r = \frac{D}{Dt}\hat{V}^r = \frac{D}{Dt}r= v^r \left(1 + r\left(\frac{1}{2}\frac{d}{dr}\ln(g_{rr})-\frac{1}{2}\frac{d}{dr}\ln(g_{rr})\right)\right) = v^r
\end{equation}
Thus, we are right about our calculations that
\begin{equation}
    \beta_r = \frac{Dr}{Dt} = \frac{dr}{dt}
\end{equation}
    \item The azimuthal component
    \begin{equation}
    \beta_\phi = \frac{D}{Dt}\hat{V}^\phi = \frac{\sqrt{g_{rr}g_{\phi\phi}}}{2r \sin^2{\theta}}\left[g_{00}^2 \frac{d}{dr}\left(\frac{g_{0\phi}}{g_{00}}\right)+\left[\beta_r\Phi'(r)+\Omega\right]\left(g_{00}\frac{dg_{\phi\phi}}{dr}+g_{0\phi}\frac{dg_{0\phi}}{dr}\right)\right]
\end{equation}
\end{enumerate}
The velocity vector can be written as (still working in the normalized basis)
\begin{equation}
    \boldsymbol{v} = \beta_r\hat{\boldsymbol{e}}_r + \beta_\phi \hat{\boldsymbol{e}}_\phi
\end{equation}
In the case of the Schwarzshild metric ($a=0$), 
\begin{equation}
    \Phi'(r) = -\frac{\Omega_1(r)}{\alpha^2}
\end{equation}
The velocity vector is (see how it reduces to the flat spacetime case in (\ref{SpeedFlatArch}) when we have Archimedean spiral, $\Omega_1(r)=\Omega$)
\begin{equation}
    {\boldsymbol{v} = \beta_r \boldsymbol{e}_r + r\Omega\left(\alpha-\frac{1}{\alpha}\frac{\Omega_1(r)}{\Omega}\beta_r\right)\boldsymbol{e}_\phi} \hspace{1cm}\underset{\alpha=1}{\overset{M=0}{\to}}\hspace{1cm}\boldsymbol{v} = \beta_r \boldsymbol{e}_r + r\Omega\left(1-\frac{\Omega_1(r)}{\Omega}\beta_r\right)\boldsymbol{e}_\phi
\end{equation}
Let us calculate the ratio generally in figure (\ref{pitchangle}) in the equatorial plane 
\begin{equation}
    \tan(\chi) = \frac{\beta_\phi}{\beta_r} = \frac{\sqrt{g_{rr}g_{\phi\phi}}}{2r}\left[\frac{g_{00}^2}{\beta_r} \frac{d}{dr}\left(\frac{g_{0\phi}}{g_{00}}\right)+\left[\Phi'(r)+\frac{\Omega}{\beta_r}\right]\left(g_{00}\frac{dg_{\phi\phi}}{dr}+g_{0\phi}\frac{dg_{0\phi}}{dr}\right)\right]
\end{equation}
Defining a decreasing (dimensionless) function $\delta(r)$ that is the change between the actual shape of the wire and the approximate form (the Archimedean spiral) relative to the latter:
\begin{equation}
    \Omega_1(r) = \Omega(1+\delta(r)) \hspace{3cm} \delta(r) \ll 1 \:\: \forall r \gtrapprox r_2 \hspace{2cm} \lim_{r\to\infty}\delta(r) = 0
\end{equation}
At very high energy ($E\to\infty$), 
\begin{equation}
    \beta_r \hspace{0.5cm}\overset{E\to\infty}{\to}\hspace{0.5cm} \frac{G_{00}}{G_{0r}+\sqrt{G_{00}G_{rr}+G_{0r}^2}} 
\end{equation}
Let us get the value of the ratio at the flat light cylinder position $r_{2;flat} = 1/\Omega$ (which is close to the Kerr one, but $r_{2;flat} > r_2$) for the Schwarzshild black hole and expand in small $\delta$
\begin{equation}
    \lim_{E\to\infty}\tan(\chi (r=1/\Omega))_{a= 0} =\frac{1}{2\eta}\left[(1+\delta)\sqrt{1-2\eta} - \sqrt{(1+\delta)^2-2\eta}\right]\approx \frac{-\delta}{\sqrt{1-2\eta}}\hspace{2cm} \eta < \frac{1}{2}
\end{equation}
In the case of the flat spacetime ($\eta\to 0)$, the ratio does not depend on the rotation at all: 
\begin{equation}
    \lim_{E\to\infty}\tan(\chi (r=1/\Omega))_{M=0} =-\frac{1}{2}\frac{\delta(\delta+2)}{\delta+1} \approx -\delta
\end{equation}
In the case of a perfect Archimedean spiral, we have a zero ratio exactly in both Schwarzschild and flat spacetimes. Generally speaking, both results are close to zero for small $\delta$. 
The same analysis can be done in the Kerr case, but we will not include the exact solution. Instead, we show graphically that even in the case of a maximally spinning black hole (which results in a greater value of the azimuthal speed, $\beta_\phi$), the ratio is still close to zero, and can be exactly zero for some variation.
}\begin{figure}[tbh!]
\centering
\includegraphics[width=1\linewidth]{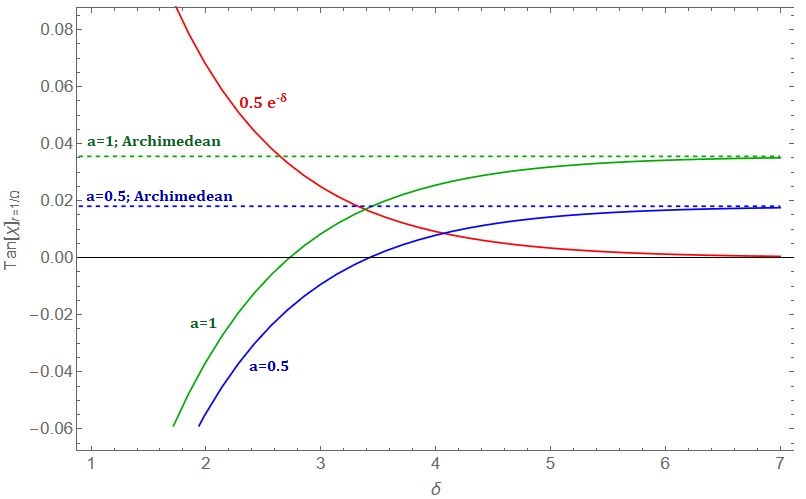}
\caption{{The ratio $\tan(\chi)=\beta_\phi/\beta_r$ (for high energy) at the outer light cylinder (in the flat case; i.e., $r=1/\Omega$) versus a free parameter $\delta$ to control the variation in the wire shape from the Archimedean case, for mass-rotation parameter $\eta=0.1$. The variation allowed range here is $\left(0,\frac{1}{2}\right]$, and presented as an exponentially decaying function of the parameter $\delta$, and is plotted in red. We have two sets of curves corresponding to two different wire shapes: the dashed lines correspond to the Archimedean spiral case, and the solid ones correspond to a spiral that is slightly different from the Archimedean one. The colors of the curves correspond to the values of the Kerr parameter: blue curves are for $a=0.5$ and green curves are for maximally spinning black hole case ($a=1$). For some variations, the curves meet the horizontal axis, meaning that the ratio is zero.}}\label{COMP2}
\end{figure}
{For plotting purposes, we consider the following possible values of the variation: $\delta(r=1/\Omega)\in\left(0, \frac{1}{2}\right]$. Moreover, we present these possible values in a form of an exponentially decaying function $e^{-\delta}/2$, where $\delta$ here is just a free parameter, not the variation. 
We choose the allowed values of $\{\eta, a\}$ based on the following constraints:
\begin{equation}
    \eta < \frac{1}{a+2\sqrt{2}}<\frac{1}{2\sqrt{2}}<\frac{1}{2}\hspace{1cm}\eta \sqrt{\frac{(1-a\eta)}{(1+a\eta)^3}}< \frac{1}{3\sqrt{3}} 
\end{equation}
We show in figure (\ref{COMP2}) that the small variation can cause the ratio in the Kerr case to approach zero at the flat-case light cylinder position. However, this might not be the case for bigger values of $\eta$; in that case, we expect the ratio to approach zero further a bit from the light cylinder (i.e., at some finite $r>1/\Omega$) because of the decaying nature of the ratio as a function of the radial coordinate $r$ in general.}

{Moreover, far away from the light cylinder, the ratio in the Kerr case becomes
\begin{equation}
    \lim_{E\to\infty}\tan(\chi (r \gg 1/\Omega)) \approx -\sqrt{\delta(\delta+2)} + \frac{1+\delta}{r\Omega} 
\end{equation}
Thus,
\begin{equation}
    \lim_{E\to\infty}\lim_{r\to\infty}\tan(\chi) = -\sqrt{\delta(\delta+2)} \approx -\sqrt{2\delta}
\end{equation}
Also, the Lorenta factor (relative to the observer at infinity) far away from the light cylinder for large energy is
\begin{equation}
    \gamma_\beta \approx \frac{1+\delta}{\sqrt{\delta(\delta+2)}} \approx \frac{1}{\sqrt{2 \delta}} \to \infty
\end{equation}}

\begin{figure}[tbh!]
\centering
\includegraphics[width=1\linewidth]{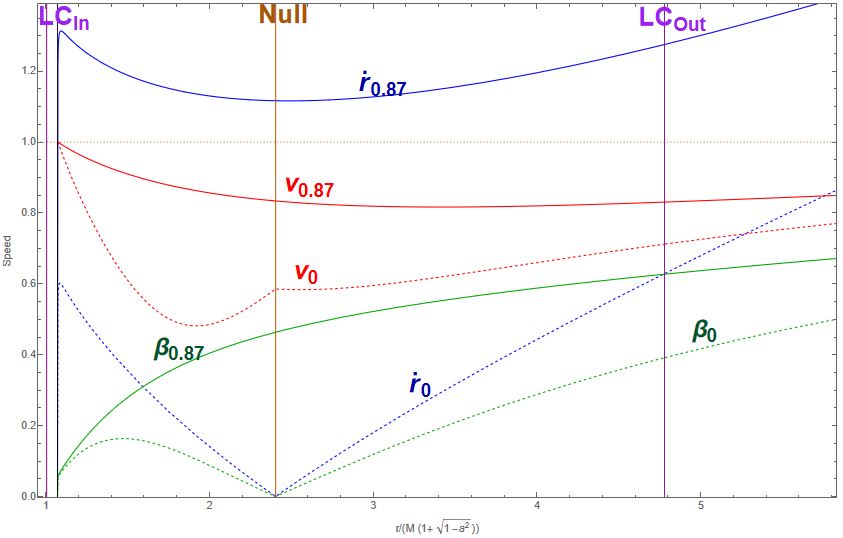}
    \caption{Various  velocities for a particle  moving in  Kerr \BH, in  the equatorial plane for  Kerr parameter $a=0.5$ and mass-rotation parameter $\eta =0.1$.  The coordinate $r$ here is taken relative to the Kerr horizon $M(1+\sqrt{1-a^2})$. We have
    two sets of curves corresponding to two different initial Lorentz factors $\gamma_{r0}$ in (\ref{ObservedVelocity}) at the null line: the dashed curves correspond to a zero initial speed ($\gamma_{r0} = 1$), and the solid ones correspond to an initial speed of 0.87 of the speed of light ($\gamma_{r0}\approx 2.03)$. The colors of the curves correspond to the different kinds of speeds: (i) blue curves are the proper speed, $\dot{r}$ in (\ref{VelocitySquared2}); (ii) green curves are speed $\beta_{r}$, the speed measured by an observer at infinity (iii)  red curves are the speed correspond to the Lorentz factor in (\ref{GammaFactorGenerallyFinal2}). The dotted (brown) horizontal line at $1$ is plotted to indicate the speed of light (we see that $dr/d\tau$ exceeds the speed of light between the two light cylinders, while the other two speeds do not). The vertical lines are colored as follows: purple for both light cylinders (the inner one is so close to the Kerr horizon at 1); brown for the null line, which lies between the light cylinders; and, black for the ergosphere (\ref{ERGOSPHEREINAPPENDIX}). The ergosphere lies before the null line due to a condition on both $\eta, \: a$, in addition to the condition (\ref{SpecificCubicEquationConstraint}): $|a-\eta^{-1}| > 2\sqrt{2}$. Finally, we only plotted the speeds outside the ergosphere ($\mathcal{R} = 2M$ here) because the Lorentz factor in (\ref{GammaFactorGenerallyFinal2}) is only defined after such radius.
}
\label{FIGUREALL}
\end{figure}

Next, we calculate the Lorentz factors relative to fixed observers in spacetime and co-rotating observers with the wire. The Lorentz factor can be used to get the speed of the particle as a function of the distance from the black hole singularity. This speed will never exceed the speed of light, unlike the proper speed in (\ref{VelocitySquared2}), see figure (\ref{FIGUREALL}). We construct the Lorentz factor in curved spacetimes by imagining filling the (2+1)-D Kerr spacetime with momentarily fixed observers everywhere (only between the ergosphere and the outer light cylinder, as we will show why later). The particle anywhere between the light cylinders is at the same location as exactly one of these observers. At this point, we can obtain a dot product between the four-velocity of the fixed observer and that of the particle, and get an invariant quantity, which is the Lorentz factor of the particle as seen by the fixed observer where the particle is. The four-velocity of a fixed observer is ($\dot{t}\geq 0, \dot{r}=0,\dot{\phi}=0$)
\begin{equation}
U^\mu_{fixed} = \begin{pmatrix}
    \dot{t} \\0\\0
\end{pmatrix} \xrightarrow[\mathcal{L}=-1]{\mathcal{L}=-g_{00}\,\dot{t}^2} U^\mu_{fixed} =\frac{1}{\sqrt{g_{00}\,}}\begin{pmatrix}
    1 \\0\\0
\end{pmatrix} \Rightarrow U_\mu^{fixed} = \begin{pmatrix}
    -\sqrt{g_{00}\,} && 0 && \frac{g_{0 \phi}\,}{\sqrt{g_{00}\,}}
\end{pmatrix}
\end{equation}
The four-velocity of the particle is
\begin{equation}
    U^\mu_{particle} = \begin{pmatrix}
    \dot{t} \\\dot{r}\\ \dot{\phi}\end{pmatrix} = \begin{pmatrix}\dot{t} \\\dot{r}\\ \Phi'(r)\dot{r}+\Omega\dot{t}\end{pmatrix}
\end{equation}
The relative Lorentz factor observed by such fixed observes for the particle is
\begin{equation}\label{GammaFactorGenerally}
    \gamma = - U_\mu^{fixed} U^\mu_{particle} = \left(\sqrt{g_{00}\,}-\frac{\Omega g_{0 \phi}\,}{\sqrt{g_{00}\,}}\right)\dot{t}-\frac{\Phi'(r) g_{0 \phi}\,}{\sqrt{g_{00}\,}}\dot{r}
\end{equation}
From both (\ref{tDOT}) and (\ref{VelocitySquared2}), the Lorentz factor becomes
\begin{multline}\label{GammaFactorGenerallyFinal}
    \gamma = \left(\sqrt{g_{00}\,}-\frac{\Omega g_{0 \phi}\,}{\sqrt{g_{00}\,}}\right)\frac{E}{G_{00}\,} \pm \\ \left(\left(\sqrt{g_{00}\,}-\frac{\Omega g_{0 \phi}\,}{\sqrt{g_{00}\,}}\right)\frac{G_{0r}\,}{G_{00}\,}-\frac{\Phi'(r) g_{0 \phi}\,}{\sqrt{g_{00}\,}}\right)\sqrt{\frac{E^2 - G_{00}\,}{ G_{0r}^2+G_{00}\,G_{rr}\, }}
\end{multline}

The negative (positive) sign means that the particle is moving towards (away from) the black hole. This Lorentz factor is only defined whenever $g_{00}$ is positive; i.e., outside the ergosphere
\begin{equation}\label{ERGOSPHEREINAPPENDIX}
    \mathcal{R} = M\left(1+\sqrt{1-a^2 \cos^2{\theta}}\right)
\end{equation}
See Fig. \ref{FIGUREALL} for a plot of the speed of the particle observed by the fixed observers, which can be obtained from the relation
\begin{equation}
    \gamma=\frac{1}{\sqrt{1-v^2}}
\end{equation}

We have a range of allowed values for the constant of motion $E$. Let $r_1$ and $r_2$ be the locations of the light cylinders. From (\ref{VelocitySquared2}), we get
\begin{equation}\label{EFIRSTConstraints}
    0 < G_{00}\, \leq E^2, \hspace{3cm} r\in (r_1, r_2)
\end{equation}
\begin{equation}
    \Omega^2 r^3 - (1-a^2\eta^2-E^2) r + 2M(1 -a\eta)^2 \geq 0
\end{equation}
This is a cubic equation on the form $\Omega^{\,2}r^3-br+d = 0$ that has to be always non-negative, for all the positive values of $r$. Therefore, we need to guarantee that the minimum (i.e., extreme) value of the cubic function is also non-negative. We obtain
\begin{equation}\label{ConstraintsOnBandD2}
    \Omega^{\,2}r^3-br+d\geq 0, \hspace{2cm} b> 0, \: d \geq \frac{1}{3\sqrt{3}}\:\frac{2 b^\frac{3}{2}}{\Omega}
\end{equation}
\begin{equation}\label{AllConstraints}
    \left\{\begin{array}{cc}
        \sqrt{1-a^2\eta^2-3\eta^\frac{2}{3}(1-a\eta)^\frac{4}{3}} \leq E &  r \in (r_1, r_2)\\
        0 \leq E < \infty & r > r_2
    \end{array}\right.
\end{equation}


The time constant $E$ that appears in (\ref{GammaFactorGenerallyFinal}) can be related to the initial conditions as observed by a set of fixed observers along the rotating wire. We imagine a set of fixed observers along the rotating wire, only defined between the two light cylinders, and they observe the particle radial speed only. 
We obtain the relative Lorentz factor as seen by the observers along the wire by applying the same procedure we used to get (\ref{GammaFactorGenerallyFinal}), but now on the (1+1)-D sub-manifold.
The four-velocity of a fixed observer is ($\dot{t}\geq 0, \dot{r}=0$)\begin{equation}
U^\mu_{fixed} = \begin{pmatrix}
    \dot{t} \\0
\end{pmatrix} \xrightarrow[\mathcal{L}=-1]{\mathcal{L}=-G_{00}\,\dot{t}^2} U^\mu_{fixed} =\frac{1}{\sqrt{G_{00}\,}}\begin{pmatrix}
    1 \\0
\end{pmatrix} \Rightarrow U_\mu^{fixed} = \begin{pmatrix}
    -\sqrt{G_{00}\,} &&\frac{G_{0r}\,}{\sqrt{G_{00}\,}}
\end{pmatrix}
\end{equation}
The four-velocity of the particle is
\begin{equation}
    U^\mu_{particle} = \begin{pmatrix}\dot{t} \\\dot{r}\end{pmatrix}\hspace{1cm}\xrightarrow[]{(\ref{tDOT})}\hspace{1cm}U^\mu_{particle} = \begin{pmatrix} 
\frac{G_{0r}\, \dot{r} + E}{G_{00}\,}\\  
\dot{r}
\end{pmatrix}
\end{equation}
The radial Lorentz factor of the radially moving particle relative to the fixed observers (on the wire) is the invariant quantity
\begin{equation}\label{ObservedVelocity}
    \gamma_r = - U_\mu^{fixed} U^\mu_{radial} = \frac{E}{\sqrt{G_{00}\,}} = \frac{1}{\sqrt{1-v_r^2}}
\end{equation}
As expected, the radial Lorentz factor coincides with the total one in (\ref{GammaFactorGenerallyFinal}) in the case of a non-rotating wire:
\begin{equation}
    \gamma = \gamma_r=\frac{E}{\sqrt{G_{00}}}=\frac{E}{\sqrt{g_{00}}} \hspace{4cm} (\Omega = 0)
\end{equation}
Since the Lorentz factor is always positive, the constant $E$ is always positive too. We can get $E$ easily from the initial conditions as seen by a fixed observer along the wire. If such an observer is at $r_0$ observes the particle moving along the wire a radial Lorentz factor $\gamma_{r0}$, we obtain\footnote{For clarity, the difference between $\gamma_{r0}$ and $\gamma$ is that the former is obtained by the observers along (and rotating with) the wire (i.e., in the (1+1)-D sub-manifold); the latter, $\gamma$, is obtained by the observers fixed on the Kerr spacetime (i.e., in the (2+1)-D manifold), which contains both the radial and the rotational parts of the speed of the particle. 
}
\begin{equation}\label{ActualValueOfTheTimeConstantE}
    E = \gamma_{r} \sqrt{G_{00}}=\gamma_{r0} \sqrt{G_{00}(r_0)}
\end{equation}
The Lorentz factor in (\ref{GammaFactorGenerallyFinal}) becomes \begin{multline}
    \gamma = \frac{E}{G_{00}}\left[ \left(\sqrt{g_{00}}-\frac{\Omega g_{0 \phi}}{\sqrt{g_{00}}}\right)\pm \frac{v_r}{\sqrt{G_{0r}^2+G_{00}\,G_{rr} }}\left(\left(\sqrt{g_{00}}-\frac{\Omega g_{0 \phi}}{\sqrt{g_{00}}}\right)G_{0r} -\frac{\Phi'(r) g_{0 \phi}}{\sqrt{g_{00}}}G_{00}\right)\right]\\ =  \frac{\gamma_r}{\sqrt{G_{00}}}\left[ \left(\sqrt{g_{00}}-\frac{\Omega g_{0 \phi}}{\sqrt{g_{00}}}\right)\pm \frac{v_r}{\sqrt{G_{0r}^2+G_{00}\,G_{rr} }}\left(\left(\sqrt{g_{00}}-\frac{\Omega g_{0 \phi}}{\sqrt{g_{00}}}\right)G_{0r} -\frac{\Phi'(r) g_{0 \phi}}{\sqrt{g_{00}}}G_{00}\right)\right]
\end{multline}
\begin{multline}\label{GammaFactorGenerallyFinal2}
    \gamma=\gamma_{r0}\:\frac{\sqrt{G_{00}(r_0)}}{G_{00}\,}\left(\sqrt{g_{00}\,}-\frac{\Omega g_{0 \phi}\,}{\sqrt{g_{00}\,}}\right) \pm \\ \left(\left(\sqrt{g_{00}\,}-\frac{\Omega g_{0 \phi}\,}{\sqrt{g_{00}\,}}\right)\frac{G_{0r}\,}{G_{00}\,}-\frac{\Phi'(r) g_{0 \phi}\,}{\sqrt{g_{00}\,}}\right)\sqrt{\frac{\gamma_{r0}^2 G_{00}(r_0)-G_{00}\,}{ G_{0r}^2+G_{00}\,G_{rr}\, }}
\end{multline}
We imagine ejecting the particles on the outer light cylinder ($r=r_2$) with an initial Lorentz factor $\gamma_{0}$ as the initial conditions. Then, we solve for the energy $E$:
\begin{equation}
    \gamma_0 = \lim_{r\to r_2} \gamma = \frac{1}{2E} \left(\sqrt{g_{00}\,}-\frac{\Omega g_{0 \phi}\,}{\sqrt{g_{00}\,}}\right)_{r_2}\left(1+G_{rr}\left(\frac{E}{G_{0r}}\right)^2\right)_{r_2} - \left(\frac{\Phi'(r) g_{0 \phi}}{\sqrt{g_{00}}} \frac{E}{|G_{0r}|}\right)_{r_2} 
\end{equation}
\begin{equation}
    E = \left\{\frac{\gamma_0  \pm \sqrt{\gamma_0^2 - \left(\sqrt{g_{00}\,}-\frac{\Omega g_{0 \phi}\,}{\sqrt{g_{00}\,}}\right) \left[\left(\sqrt{g_{00}\,}-\frac{\Omega g_{0 \phi}\,}{\sqrt{g_{00}\,}}\right)\frac{ G_{rr}}{G_{0r}^2} - \frac{2}{|G_{0r}|}\frac{\Phi'(r) g_{0 \phi}}{\sqrt{g_{00}}}\right]}}{\left(\sqrt{g_{00}\,}-\frac{\Omega g_{0 \phi}\,}{\sqrt{g_{00}\,}}\right)\frac{ G_{rr}}{G_{0r}^2} - \frac{2}{|G_{0r}|}\frac{\Phi'(r) g_{0 \phi}}{\sqrt{g_{00}}}}\right\}_{r_2}
\end{equation}
Therefore, we see that there has to be a minimum initial Lorentz factor for the constant $E$ to have real values. Moreover, we only need the larger energy of the two solutions to guarantee that we have a solution between the light cylinders as well, by being consistent with the constraints (\ref{AllConstraints}). 
Usually, we have a large Lorentz factor when the particle reaches the outer light cylinder; thus, giving the particle an initial push at the outer light cylinder with a relatively large $\gamma_0$ will result in a large amount of the energy:
\begin{equation}
    E \approx \gamma_0\left(\frac{2 G_{0r}^2}{\left(\sqrt{g_{00}\,}-\frac{\Omega g_{0 \phi}\,}{\sqrt{g_{00}\,}}\right)G_{rr}-2 |G_{0r}| \frac{\Phi'(r) g_{0 \phi}}{\sqrt{g_{00}}}}\right)_{r_2} \sim \gamma_0
\end{equation}
For large values of the Lorentz factor of the particle at the outer light cylinder, the energy of the particle behaves the same way as the Lorentz factor of the particle at the outer light cylinder.

 \end{document}